\documentclass[submission,copyright,creativecommons]{eptcs}
\providecommand{\event}{SAIRP 2013}
\usepackage[utf8]{inputenc}
\usepackage[T1]{fontenc}

\usepackage{tikz}
\usepackage{subcaption}
\usepackage{array}
\usepackage{multirow}
\usepackage{proof}
\usepackage{amsfonts, amssymb}
\usepackage{amsmath}
\usepackage{placeins}

\usepackage{mathpartir}

\usepackage{xifthen}

\def\Val{\ensuremath{\mathit{Val}}}
\def\Loc{\ensuremath{\mathit{Loc}}}
\def\ALoc{\ensuremath{\mathit{ALoc}}}
\def\Env{\ensuremath{\mathit{Env}}}
\def\Heap{\ensuremath{\mathit{Heap}}}
\def\Var{\ensuremath{\mathit{Var}}}
\def\PP{\ensuremath{\mathit{PP}}} 
\def\Field{\ensuremath{\mathit{Field}}}
\def\Trace{\ensuremath{\mathit{Trace}}}
\def\Dep{\ensuremath{\mathit{Dep}}}
\def\Stor{\ensuremath{\mathit{Store}}}
\def\Source{\ensuremath{\mathit{Source}}}

\newcommand{\abs}[1]{{#1}^\sharp}

\makeatletter
\let \mpr@@andcr \mpr@andcr
\def \mpr@andcr {\futurelet \mpr@next \mpr@andcr@}
\def \mpr@andcr@ {\ifx \mpr@next \noindent \mpr@par\else \mpr@@andcr\fi}
\def \MathparBindings {\let \mpr@par \par \mpr@bindings}
\makeatother

\usepackage{xspace}

\usepackage{listings}
\usepackage{lstcoq}

\tikzstyle{every picture}+=[remember picture]

\tikzset{
	location/.style = {
		draw = Aluminium6,
		rectangle,
		rounded corners = 2pt,
		text centered,
		minimum height = 5mm,
		minimum width = 10mm,
		top color = white,
		bottom color = Aluminium1
	},
	new location/.style = {
		location,
		top color = LightOrange,
		bottom color = Orange
	},
	red location/.style = {
		location,
		top color = Aluminium1,
		bottom color = LightScarletRed
	}
}

\lstdefinelanguage{JavaScript}{
  keywords={typeof, new, true, false, try, catch, finally, function, return, catch, switch, var, if, in, while, for, do, else, case, break, continue, with, delete},
  keywordstyle=\color{DarkSkyBlue}\bfseries,
  ndkeywords={class, export, boolean, throw, implements, import, this, undefined, null, NaN, eval},
  ndkeywordstyle=\color{Plum}\bfseries,
  identifierstyle=\color{Aluminium5},
  sensitive=false,
  comment=[l]{//},
  morecomment=[s]{/*}{*/},
  commentstyle=\color{Chameleon}\ttfamily,
  stringstyle=\color{Orange}\ttfamily,
  morestring=[b]',
  morestring=[b]"
}

\definecolor{LightButter}{rgb}{0.98,0.91,0.31}
\definecolor{LightOrange}{rgb}{0.98,0.68,0.24}
\definecolor{LightChocolate}{rgb}{0.91,0.72,0.43}
\definecolor{LightChameleon}{rgb}{0.54,0.88,0.20}
\definecolor{LightSkyBlue}{rgb}{0.45,0.62,0.81}
\definecolor{LightPlum}{rgb}{0.68,0.50,0.66}
\definecolor{LightScarletRed}{rgb}{0.93,0.16,0.16}
\definecolor{Butter}{rgb}{0.93,0.86,0.25}
\definecolor{Orange}{rgb}{0.96,0.47,0.00}
\definecolor{Chocolate}{rgb}{0.75,0.49,0.07}
\definecolor{Chameleon}{rgb}{0.45,0.82,0.09}
\definecolor{SkyBlue}{rgb}{0.20,0.39,0.64}
\definecolor{Plum}{rgb}{0.46,0.31,0.48}
\definecolor{ScarletRed}{rgb}{0.80,0.00,0.00}
\definecolor{DarkButter}{rgb}{0.77,0.62,0.00}
\definecolor{DarkOrange}{rgb}{0.80,0.36,0.00}
\definecolor{DarkChocolate}{rgb}{0.56,0.35,0.01}
\definecolor{DarkChameleon}{rgb}{0.30,0.60,0.02}
\definecolor{DarkSkyBlue}{rgb}{0.12,0.29,0.53}
\definecolor{DarkPlum}{rgb}{0.36,0.21,0.40}
\definecolor{DarkScarletRed}{rgb}{0.64,0.00,0.00}
\definecolor{Aluminium1}{rgb}{0.93,0.93,0.92}
\definecolor{Aluminium2}{rgb}{0.82,0.84,0.81}
\definecolor{Aluminium3}{rgb}{0.73,0.74,0.71}
\definecolor{Aluminium4}{rgb}{0.53,0.54,0.52}
\definecolor{Aluminium5}{rgb}{0.33,0.34,0.32}
\definecolor{Aluminium6}{rgb}{0.18,0.20,0.21}

\definecolor{darkGreen}{rgb}{0,0.5,0}

\usetikzlibrary{decorations.pathmorphing}
\pgfdeclarelayer{background}
\pgfdeclarelayer{foreground}
\pgfsetlayers{background,main,foreground}

\newtheorem{prop}{Property}

\newcommand\reducebot[1][]{\ifthenelse{\equal{#1}{}}{\Downarrow}{\Botarrow_{#1}}}

\newcommand\javascript{\textsc{JavaScript}\xspace}
\newcommand\whilel{\textsc{While}\xspace}

\newcommand\coqn{\textsc{Coq}\xspace}
\newcommand\caml{\textsc{OCaml}\xspace}

\newcommand\nanoJS{\textsc{O'While}\xspace}

\xdef\isgreyexpr{n}
\newcommand\grey[1]{%
	\def\greyargument{#1}%
	\ifthenelse{\equal\isgreyexpr{n}}{
		\xdef\isgreyexpr{y}%
		\textcolor{Aluminium4}{\greyargument}\,%
		\xdef\isgreyexpr{n}%
		}{\greyargument}%
}

\xdef\isinunderlinejsexpr{n}
\newcommand\jsexpr[2][Aluminium3]{%
	\ifthenelse{\equal\isgreyexpr{n}}{\textcolor{#1}}{}{%
		\def\jsexprargument{\ifthenelse{\equal\isgreyexpr{n}}{\textcolor{black}}{}{#2}}%
		\ifthenelse{\equal\isinunderlinejsexpr{n}}{
			\xdef\isinunderlinejsexpr{y}%
			\underline\jsexprargument%
			\xdef\isinunderlinejsexpr{n}%
			}{\jsexprargument}%
}}

\newcommand\ifthenabove[1]{%
	\ifthenelse{\equal{#1}{}}{}{^{#1}}%
}

\newcommand\aloc[2]{#1^{#2}}

\newcommand\jscom[1]{\textrm{\lstinline|#1|}}

\newcommand\jsid[2][]{{\textup{\texttt{\small #2}}\ifthenabove{#1}}}

\def\nsem{\,;_{\scalebox{.5}1}\xspace}
\def\nseq{\,=_{\scalebox{.5}1}\xspace}
\def\nseqq{\,=_{\scalebox{.5}2}\xspace}
\def\nsop{\;\texttt{op}_{\scalebox{.5}1}\;}
\def\nsopp{\;\texttt{op}_{\scalebox{.5}2}\;}
\def\nsdot{\,._{\scalebox{.5}1}\xspace}

\newcommand\seqq[2]{\jsexpr{#1 \nsem #2}}

\newcommand\jsif[3]{\jsexpr{\jscom{if}~#1~\jscom{then}~#2~\jscom{else}~#3}}
\newcommand\jsiff[3]{\jsexpr{\jscom{if1}(#1,#2,#3)}}

\newcommand\jswhile[2]{\jsexpr{\jscom{while}~#1~\jscom{do}~#2}}
\newcommand\jswhilee[3]{\jsexpr{\jscom{while1}(#1,#2,#3)}}
\newcommand\jswhileee[3]{\jsexpr{\jscom{while2}(#1,#2,#3)}}

\newcommand\jsass[3][]{\jsexpr{\jsid{#2}\ifthenabove{#1}~\jscom{=}~#3}}
\newcommand\jsasss[3][]{\jsexpr{\jsid{#2}\ifthenabove{#1} \nseq #3}}

\newcommand\jsdelete[3][]{\jsexpr{\jscom{delete}~{#2}\ifthenabove{#1}\jscom{.}\jsid{#3}}}
\newcommand\jsdeletee[3][]{\jsexpr{\jscom{delete1}\ifthenabove{#1}~#2\jscom{.}\jsid{#3}}}

\newcommand\jsfieldass[4][]{\jsexpr{#2.\jsid{#3}\ifthenabove{#1}~\jscom{=}~#4}}
\newcommand\jsfieldasss[3]{\jsexpr{#1.\jsid{#2} \nseq #3}}
\newcommand\jsfieldassss[3]{\jsexpr{#1.\jsid{#2} \nseqq #3}}

\newcommand\jsop[2]{\jsexpr{#1~\texttt{op}~#2}}
\newcommand\jsopp[2]{\jsexpr{#1 \nsop #2}}
\newcommand\jsoppp[2]{\jsexpr{#1 \nsopp #2}}

\newcommand\jsobj[2][]{\jsexpr{\jscom\{#2\jscom\}\ifthenabove{#1}}}

\newcommand\jsfield[3][]{\jsexpr{#2\jscom{.}\jsid[#1]{#3}}}
\newcommand\locfield[3][]{#2\jscom{.}\jsid{#3}\ifthenabove{#1}}
\newcommand\jsfieldd[2]{\jsexpr{#1\nsdot\jsid{#2}}}

\newcommand\trace[1]{[\small{#1}]}
\newcommand\irule[1]{\textsc{#1}_i}
\newcommand\orule[1]{\textsc{#1}_o}

\newcommand\mset[1]{\ifthenelse{\isempty{#1}}%
	{\emptyset}%
	{\left\{%
		#1%
	\right\}}%
}

\newcommand\absto{\,\abs\to\,}
\newcommand\absset\mset
\newcommand\absheap\absset

\newsavebox\bbracksb
\newlength\bbracklength

\newcommand\snakeArrow[1][]{%
	\tikz [baseline, ->, #1]
		\draw [decorate, decoration = {snake,
			amplitude = .4mm, segment length = 2mm, post length = .5mm},
			#1] (0pt, 0.5ex) -- (1em, 0.5ex);}

\newcommand\flowitem{\,%
	\setbox0\hbox{$\snakeArrow$}%
	\rlap{\hbox to \wd0{\hss$\subset$\hss}}\box0%
\,}

\newcommand\powerset[1][]{{\mathcal{P}%
	\ifthenelse{\equal{#1}{}}{}{\left(#1\right)}%
}}

\newcommand\concat{\ensuremath{+\!\!\!\!\!+\,}}

\newcommand\audn\textit

\newcommand\objatpp[1]{%
	o^{#1}%
}

\newcommand{\res}{r}
\newcommand{\err}{\mathtt{err}}
\newcommand\abortc[1]{\mathtt{abort}(#1)}

\newcommand\st[1]{\mathtt{st}(#1)}

\title{Pretty-big-step-semantics-based \\ Certified Abstract Interpretation \\ (Preliminary version)}
\author{Martin Bodin  \institute{ENS Lyon and Inria} \and Thomas Jensen \institute{Inria} \and Alan Schmitt \institute{Inria}}

\def\titlerunning{Pretty-big-step-semantics-based Certified Abstract Interpretation}
\def\authorrunning{M. Bodin, T. Jensen, \& A. Schmitt}

\begin{document}
\maketitle

\begin{abstract}
We present a technique for deriving semantic program analyses from a natural semantics specification of the
programming language. The technique is based  on a
particular kind of semantics called pretty-big-step semantics. We
present a pretty-big-step semantics of a language with simple objects
called O'While and specify a series of instrumentations of the
semantics that explicitates the flows of values in a program. This
leads to  a semantics-based dependency analysis, at the core, \emph{e.g.}, 
of tainting analysis in software security. The formalization
has been realized with the Coq proof assistant.
\end{abstract}

\section{Introduction}

David Schmidt gave an invited talk at the 1995 Static Analysis
Symposium~\cite{Schmidt:95:Natural} in which he argued for using natural semantics as a
foundation for designing semantic program analyses within the
abstract interpretation framework. With natural (or ``big-step'' or ``evaluation'')
semantics, we can indeed hope to benefit from the compositional
nature of a denotational-style semantics while at the same time being
able to capture intensional properties that are best expressed using
an operational semantics.
Schmidt showed how a control flow analysis of a core higher-order
functional language can be expressed elegantly in his
framework. Subsequent work by Gouranton and Le M\'etayer showed how
this approach could be used to provide a natural semantics-based
foundation for program slicing~\cite{Gouranton:99:Dynamic}. 

In this paper, we will pursue the research agenda set out by Schmidt
and investigate further the systematic design
of semantics-based program analyses based on big-step semantics.
Two important issues here will be those of scalability and
mechanization.
The approach worked nicely for a language whose semantics could be
defined in 8 inference rules. How will it react when applied to
full-blown languages where the semantic definition comprises
hundreds of rules? Strongly linked to this question is that of how the
framework can be mechanized and put to work on larger languages using
automated tool support. In the present work, we investigate how the Coq proof assistant
can serve as a tool for manipulating the semantic definitions and certifying the
correctness of the derived static analyses.

Certified static analysis is concerned with developing static
analyzers inside proof assistants with the aim of producing a static
analyzer and a machine-verifiable proof of its semantic correctness.
One long-term goal of the work reported here is to be able to provide
a mechanically verified static analysis for the full \javascript language based on
the Coq formalization developed in the JSCert project~\cite{JSCert}.
\javascript, with its rich but also sometimes quirky semantics, is indeed
a good \textit{raison d'\^etre} for studying certified static
analysis, in order to ensure that all of the cases in the semantics
are catered for.

In our development, we shall take advantage of some recent
developments in the theory of operational semantics. In particular, we
will be using a particular format of natural semantics call
``pretty-big-step'' semantics \cite{Chargueraud:13:Pretty} which is a
streamlined form of operational semantics retaining the format of
natural semantics while being closer to small-step operational
semantics. We will give a high-level introduction to the main features
of pretty-big-step semantics in Section~\ref{sec:pretty}

Even though it is our ultimate goal, \javascript is far too big to begin with as a goal for analysis:  its pretty-big-step semantics contains more than half a thousand rules!
We will thus start by studying a much simpler language, called
\nanoJS, which is basically a \whilel language with simple objects in
the form of extensible records.
This language is quite far from \javascript, but is big enough to
catch some issues of the analyses of \javascript objects. We present
the language and its pretty-big-step semantics in
Section~\ref{sec:language}.

To test the applicability of the approach to defining static analyses, we have
chosen to formalize a data flow dependency analysis as used \emph{e.g.}, in
tainting \cite{Schwartz:10:DynamicTaint} or ``direct information-flow'' analyses of
JavaScript~\cite{Vogt:07:Taint,Guarnieri:11:Saving}. The property we ensure is
defined in Section~\ref{sec:instrumentation} and the analysis itself is defined
in Sections~\ref{sec:analysis}.

As stated above, the scalability of the approach relies on the
mechanization that will enable the developer of the analyses to prove
the correctness of analyses with respect to the semantics, and to
extract an executable analyzer. We will show how the Coq proof assistant has
been used successfully to achieve these objectives as we go along.

\section{Pretty-Big-Step Operational Semantics}
\label{sec:pretty}

As big-step semantics, pretty-big-step semantics directly relates
terms to their results. However, pretty-big-step semantics avoids the
duplication associated with big-step semantics when features such as
exceptions and divergence are added. Since duplication in the
definitions often leads to duplication in the formalization and in the
proofs, an approach based on a pretty-big-step semantics allows to
deal with programming languages with many complex constructs. (We
refer the reader to Chargu\'{e}raud's work on pretty-big-step
semantics \cite{Chargueraud:13:Pretty} for detailed information about
this duplication.) Even though the language considered here is not
complex, we have been using pretty-big-step semantics exclusively
for our \javascript developments, thus we will pursue this approach in the
present study. 

We give an intuition on how pretty-big-step semantics works through a simple
example: the execution of a while loop. In a big-step semantics, the while loop
has one or three premises. First, the condition is evaluated. Then, in the case
it returned \jscom{true}, the statement and the rest of the loop is evaluated.
The evaluation of terms returns either a pair of a state and a value (when
evaluating an expression) or simply a state (when evaluating a
statement). Writing $S$ for states, we thus have the following.

\begin{mathpar}
    \inferrule{
      S, \jsexpr{e} \to S', \jscom{false}
    }{
      S, \jswhile{e}{s} \to S'
    } \and
    \inferrule{
      S, \jsexpr{e} \to S', \jscom{true}\\
      S', s \to S''\\
      S'', \jswhile{e}{s} \to S'''
    }{
      S, \jswhile{e}{s} \to S'''
    }
\end{mathpar}

In the pretty-big-step approach, only one sub-term is evaluated in
each rule. The result of the evaluation is gathered, along with
the remaining sub-terms, in a new syntactic construct called an
\emph{extended term}. For instance, the first reduction for the
\jscom{while} loop is as follows.

\begin{mathpar}
    \inferrule{
      S, \jsexpr{e} \to S',v\\
      S, \jswhilee{S',v}{e}{s} \to S''
    }{
      S, \jswhile{e}{s} \to S''
    }
\end{mathpar}

The $\jswhilee{S',v}{e}{s}$ term includes the result of evaluating
$\jsexpr{e}$ (namely $S',v$), as well as the information
required to evaluate the rest of the loop. The $\jswhilee{\cdot, \cdot}{\cdot}{\cdot}$ term is
evaluated using one of the two rules below. If $v$ is \jscom{false},
then the evaluation immediately returns. Otherwise a second extended
term is used. Note that the state $S$ in the conclusion of the rules is
not used, as the starting state is present in the extended term. We
still include it to ensure all the rules have the same shape.

\begin{mathpar}
    \inferrule{ }{S, \jswhilee{S',\jscom{false}}{e}{s} \to S'}
    \and
    \inferrule{
      S', s \to S''\\
      S', \jswhileee{S''}{e}{s} \to S'''
    }{
      S, \jswhilee{S',\jscom{true}}{e}{s} \to S'''
    }
\end{mathpar}

In this second case, the result of evaluating the statement
$\jsexpr{s}$, namely a new state $S''$, is stored in the extended term
$\jswhileee{S''}{e}{s}$. Finally, this new state is used to evaluate
the next iteration of the loop.

\begin{mathpar}
    \inferrule{
      S'', \jswhile{e}{s} \to S'''
    }{
      S', \jswhileee{S''}{e}{s} \to S'''
    }
\end{mathpar}

Putting it all together, here is a full derivation of one run of a loop.

\begin{mathpar}
  \inferrule{
    \inferrule{\ldots}{S, \jsexpr{e} \to S',\jscom{true}}\\
    \inferrule{
      \inferrule{\ldots}{S', s \to S''}\\
      \inferrule{
        \inferrule{\ldots}{S'', \jswhile{e}{s} \to S'''}
      }{
        S', \jswhileee{S''}{e}{s} \to S'''
      }
    }{
      S, \jswhilee{S',\jscom{true}}{e}{s} \to S'''
    }
  }{
    S, \jswhile{e}{s} \to S'''
  }
\end{mathpar}

\begin{figure}
   \centering
 \parbox[t]{0.4\textwidth}{
  \begin{align*}
	s &::=\\
	&|~ \jsexpr{\jscom{skip}} \\
	&|~ \jsexpr{s_1 ; s_2} \\
	&|~ \jsif{e}{s_1}{s_2} \\
	&|~ \jswhile{e}{s} \\
	&|~ \jsass{x}{e}\\
	&|~ \jsfieldass{e_1}{f}{e_2}\\
	&|~ \jsdelete{e}{f}
\end{align*}}\quad
\parbox[t]{0.4\textwidth}{
\begin{align*}
	e &::=\\
	&|~ \jsexpr{c} \\
	&|~ \jsexpr{\jsid{x}} \\
	&|~ \jsop{e_1}{e_2} \\
	&|~ \jsexpr{\jsobj{}} \\
	&|~ \jsfield{e}{f}
\end{align*}}
  \caption{\nanoJS Syntax}
  \label{fig:nanosyntax}
\end{figure}

\section{\nanoJS and its Pretty Big Step Semantics}
\label{sec:language}

The syntax of \nanoJS is presented in Figure~\ref{fig:nanosyntax}.
Two new constructions have been added to the syntax of expressions for
the usual \whilel language: \jsexpr{\jsobj{}} creates a new object,
and $\jsfield{e}{f}$ accesses a field of an object.  Regarding
statements, we allow the addition or the modification of a field to an
object using $\jsfieldass{e_1}{f}{e_2}$, and the deletion of the field
of an object using $\jsdelete{e}{f}$. In the following we write $t$ for terms,
\emph{i.e.} both expressions and statements.

\begin{figure}[ht]
  \centering
\parbox[t]{0.2\textwidth}{
\begin{align*}
	v &::=\\
	&|~ c \\
	&|~ l
\end{align*}}\quad
\parbox[t]{0.2\textwidth}{
\begin{align*}
	\res &::=\\
	&|~ S\\
  &|~ S,v\\
	&|~ S,\err
\end{align*}}\quad
\parbox[t]{0.2\textwidth}{
  \begin{align*}
    s_e &::= \\
	&|~ s\\
    &|~ \seqq{r}{s} \\
    &|~ \jsiff{r}{s_1}{s_2} \\
    &|~ \jswhilee{r}{e}{s} \\
    &|~ \jswhileee{r}{e}{s} \\
    &|~ \jsasss{x}{r} \\
    &|~ \jsfieldasss{r}{f}{e} \\
    &|~ \jsfieldassss{l}{f}{r} \\
    &|~ \jsdeletee{r}{f}
  \end{align*}}\quad
\parbox[t]{0.2\textwidth}{
  \begin{align*}
    e_e &::= \\
	&|~ e\\
    &|~ \jsopp{r}{e} \\
    &|~ \jsoppp{v}{r} \\
    &|~ \jsfieldd{r}{f}
  \end{align*}}
  \caption{\nanoJS Values and Extended Syntax}
  \label{fig:nanoextsyntax}
\end{figure}

Objects are passed by reference, thus values $v$ (Figure
\ref{fig:nanoextsyntax}) are either locations $l$ or primitive values
$c$. In this work, we only consider boolean primitive values.

The \emph{state} of a program contains both an \emph{environment} $E$, which is
a mapping from variables to values, and a \emph{heap} $H$, which is a mapping
from locations to objects, that are themselves mappings from fields to
values. In the following, we write $S$ for $E,H$ when there is no need to access
the environment nor the heap. Results $\res$ are either a state $S$, a pair of a
state and value $S,v$, or a pair of an error and a state $S,\err$.

\begin{figure}
  \centering
  \begin{mathpar}
    \inferrule*[right=Skip]{ }{S, \jsexpr{\jscom{skip}} \to S} \and
    \inferrule*[right=Seq]{
      S, \jsexpr{s_1} \to r \\
      S, \seqq{r}{s_2} \to r'
    }{
      S, \jsexpr{s_1; s_2} \to r'
    } \and
    \inferrule*[right=Seq1]{
      S', s \to r
    }{
      S, \seqq{S'}{s} \to r
    } \and
    \inferrule*[right=If]{
      S, \jsexpr{e} \to r \\
      S, \jsiff{r}{s_1}{s_2} \to r'
    }{
      S, \jsif{e}{s_1}{s_2} \to r'
    } \and
    \inferrule*[right=IfTrue]{
      S', s_1 \to r
    }{
      S,\jsiff{(S',\jscom{true})}{s_1}{s_2} \to r
    } \and
    \inferrule*[right=IfFalse]{
      S', s_2 \to r
    }{
      S,\jsiff{(S',\jscom{false})}{s_1}{s_2} \to r
    } \and
    \inferrule*[right=While]{
      S, e \to r\\
      S, \jswhilee{r}{e}{s} \to r'
    }{
      S, \jswhile{e}{s} \to r'
    } \and
    \inferrule*[right=WhileTrue1]{
      S', s \to r\\
      S', \jswhileee{r}{e}{s} \to r'
    }{
      S, \jswhilee{(S',\jscom{true})}{e}{s} \to r'
    } \and
    \inferrule*[right=WhileTrue2]{
      S', \jswhile{e}{s} \to r
    }{
      S, \jswhileee{S'}{e}{s} \to r
    } \and
    \inferrule*[right=WhileFalse]{ }{S, \jswhilee{(S',\jscom{false})}{e}{s} \to S'}
    \and
    \inferrule*[right=Asg]{
      E, H, e \to r\\
      E,H, \jsasss{x}{r} \to r'
    }{
      E, H, \jsass{x}{e} \to r'
    }
    \and
    \inferrule*[right=Asg1]{
      E' = E[\jsid{x} \mapsto v]
    }{
      S, \jsasss{x}{(E,H,v)} \to E', H
    }
    \and
    \inferrule*[right=FldAsg]{
      S, \jsexpr{e_1} \to r\\
      S, \jsfieldasss{r}{f}{e_2} \to r'
    }{
      S, \jsfieldass{e_1}{f}{e_2} \to r'
    }
    \and
    \inferrule*[right=FldAsg1]{
      S', e \to r\\
      S', \jsfieldassss{l}{f}{r} \to r'
    }{
      S, \jsfieldasss{(S',l)}{f}{e} \to r'
    }
    \and
    \inferrule*[right=FldAsg2]{
      H[l] = o\\
      o' = o\left[\jsid{f} \mapsto v\right]\\
      H' = H\left[l \mapsto o'\right]
    }{
      S, \jsfieldassss{l}{f}{(E,H,v)} \to E, H'
    }
	\and
    \inferrule*[right=Del]{
      S, \jsexpr{e} \to r\\
      S, \jsdeletee{r}{f} \to r'
    }{
      S, \jsdelete{e}{f} \to r'
    }
    \and
    \inferrule*[right=Del1]{
      H[l] = o\\
      o[\jsid{f}] \neq \bot\\
      o' = o\left[\jsid{f} \mapsto \bot\right]\\
      H' = H\left[l \mapsto o'\right]
    }{
      S, \jsdeletee{(E,H,l)}{f} \to E, H'
    }
    \and
    \inferrule*[right=Cst]{ }{ S, \jsexpr{c} \to S, c} \and
    \inferrule*[right=Var]{E[\jsexpr{\jsid{x}}] = v}{E, H, \jsexpr{\jsid{x}} \to E, H, v} \and
    \inferrule*[right=Bin]{
      S, \jsexpr{e_1} \to r \\
      S, \jsopp{r}{e_2} \to r'\\
    }{
      S, \jsop{e_1}{e_2} \to r'
    }
    \and
    \inferrule*[right=Bin1]{
      S', \jsexpr{e_2} \to r\\
      S', \jsoppp{v_1}{r} \to r'
    }{
      S, \jsopp{(S', v_1)}{e_2} \to r'
    }
    \and
    \inferrule*[right=Bin2]{
      v = v_1~\texttt{op}~v_2
    }{
      S, \jsoppp{v_1}{(S, v_2)} \to S,v
    }
    \and
    \inferrule*[right=Obj]{H[l] = \bot \\ H' = H[l \mapsto \jsobj{}]}
    { E, H, \jsexpr{\jsobj{}} \to E, H', l} \and
    \inferrule*[right=Fld]{
      S, \jsexpr{e} \to r \\
      S, \jsfieldd{r}{f} \to r'
    }{
      S, \jsfield{e}{f} \to r'
    }\and
    \inferrule*[right=Fld1]{H'[l] = o \\ o[\jsid{f}] = v}
    {E, H, \jsfieldd{(E',H',l)}{f} \to E', H', v}
    \and
    \inferrule*[right=Abort]{\abortc{t_e} = r}{S, t_e \to r}
  \end{mathpar}
  \caption{\nanoJS's Semantics}
  \label{fig:nanosemantics}
\end{figure}


Figure \ref{fig:nanoextsyntax} also introduces \emph{extended statements} and
\emph{extended expressions} that are used in \nanoJS's pretty-big-step
semantics, presented in Figure~\ref{fig:nanosemantics}. Extended terms $t_e$
comprise extended statements and expressions.  Reduction rules have the form $S,
t_e \to r$. The result $r$ can be an error $S',\err$. Otherwise, if $t_e$ is an
extended statement, then $r$ is a state $S'$, and if $t_e$ is an extended
expression, then $r$ is a pair of a state $S'$ and returned value $v$. We write
$\st{r}$ for the state $S$ in a result $r$.

Most rules are the usual \whilel ones, with the exception that they are given in
pretty-big-step style. We now detail the new rules for expressions and
statements. Rule \textsc{Obj} associates an empty object to a fresh location in
the heap. Rule \textsc{Fld} for the expression $\jsfield{e}{f}$ first evaluates
$e$ to some result $r$, then calls the rule for the extended expression
$\jsfieldd{r}{f}$. The rule for this extended expression is only defined if $r$
is of the form $E',H',l$ where $l$ is a location in $H'$ that points to an
object $o$ containing a field $\jsid{f}$.  The rules for field assignment and field
deletion are similar: we first evaluate the expression that defines the object
to be modified, and in the case it actually is a location, we modify this object
using an extended statement.

Finally, our semantics is parameterized by a partial function $\abortc{\cdot}$
from extended terms to results, that indicates when an error is to be raised or
propagated. More precisely, the function $\abortc{t_e}$ is defined at least if
$t_e$ is an extended term containing a subterm equal to $S,\err$ for some $S$.
In this case $\abortc{t_e} = S,\err$. We can then extend this function to define
erroneous cases. For instance, we could say that
$\abortc{\jsfieldasss{(E,H,v)}{f}{e}} = E,H,\err$ if $v$ is not a location, or
if $v = l$ but $l$ is not in the domain of $H$, or if $\jsid{f}$ is not in the domain
of $H[l]$. This function is used in the \textsc{Abort} rule, that defines when
an error is raised or propagated. This illustrates the benefit of a
pretty-big-step semantics: a single rule covers every possible error propagation
case.

The derivation in Figure \ref{fig:pretty_example} is an example of a
derivation of the semantics. It will be the basis for the running
example of this paper.

\begin{figure}
  \centering
  \begin{mathpar}
    \inferrule*[Right=Seq]{
      \inferrule*[leftskip=5em,Left=Asg]{
        \inferrule*[Left=Obj]{
          H[l] = \bot \\ H' = H[l \mapsto \jsobj{}]
        }
        {
          E, H, \jsexpr{\jsobj{}} \to E, H', l
        }\\
        \inferrule*[Right=Asg1]{
          E' = E[\jsid{x} \mapsto l]
        }{
          E,H, \jsasss{x}{E,H',l} \to E',H'
        }
      }{
        E, H, \jsass{x}{\jsobj{}} \to E', H'
      }\\
      \inferrule*[Right=Seq1,leftskip=20em,rightskip=4em,vdots=6em]{
        \inferrule*[Right=Seq]
        {
          \inferrule*[leftskip=10em,rightskip=-2em,Left=FldAsg]{
            \inferrule*[rightskip=7em,Left=Var]{
              E'[\jsid{x}] = l
            }{
              E', H', \jsexpr{\jsid{x}} \to E',H',l
            }\\
            \inferrule*[leftskip=10em,rightskip=0em,vdots=4em,Right=FldAsg1]{
              \inferrule*[Left=Obj]{
                {\begin{gathered}
                    H'[l'] = \bot \\
                    H'' = H'[l' \mapsto \jsobj{}]
                  \end{gathered}}
              }
              {
                E', H', \jsexpr{\jsobj{}} \to E', H'', l'
              }\\
              \inferrule*[Right=FldAsg2]{
                {\begin{gathered}[b]
                    H'[l] = \jsobj{}\\
                    o' = \jsobj{\jsid{f} \mapsto l'}\\
                    H''' = H''\left[l \mapsto o'\right]\end{gathered}}
              }
              {
                E', H', \jsfieldassss{l}{f}{(E', H'', l')} \to E', H'''
              }
            }
            {
              E',H',\jsfieldasss{(E',H',l)}{f}{\jsobj{}} \to E',H'''
            }
          }{
            E',H',\jsfieldass{\jsid{x}}{f}{\jsobj{}} \to E',H'''
          }\\
          \inferrule*[Right=Seq1,leftskip=25em,rightskip=13em,vdots=15em]{
            \inferrule*[Right=If]{
              \inferrule*[Left=Cst]{ }{E',H''',\jsexpr{\jscom{false}} \to E',H''',\jscom{false}}\\
              \inferrule*[Right=IfFalse,leftskip=18em,rightskip=0em,vdots=3em]{
                \inferrule*[Right=Asg]{
                  \inferrule*[Left=Obj]{
                    H'''[l''] = \bot \\ H_f = H'''[l'' \mapsto \jsobj{}]
                  }
                  {
                    E', H''', \jsexpr{\jsobj{}} \to E', H_f, l''
                  }\\
                  \inferrule*[Right=Asg1]{E_f = E'[\jsid{x} \mapsto l'']}
                  {E', H''', \jsasss{x}{(E',H_f,l'')} \to E_f,H_f}
                }{
                  E',H''',\jsass{y}{\jsobj{}} \to E_f,H_f
                }
              }{
                E',H''',\jsiff{E',H''',\jscom{false}}{\jsass{y}{\jsfield{\jsid{x}}{f}}}{\jsass{y}{\jsobj{}}} \to E_f,H_f
              }
            }
            {
              E',H''',\jsif{\jscom{false}}{\jsass{y}{\jsfield{\jsid{x}}{f}}}{\jsass{y}{\jsobj{}}} \to E_f,H_f
            }
          }{
            E', H', \seqq{(E',H''')}{\jsif{\jscom{false}}{\jsass{y}{\jsfield{\jsid{x}}{f}}}{\jsass{y}{\jsobj{}}}} \to E_f,H_f
          }
        }
        {
          E',H',\jsexpr{\jsfieldass{\jsid{x}}{f}{\jsobj{}}; \jsif{\jscom{false}}{\jsass{y}{\jsfield{\jsid{x}}{f}}}{\jsass{y}{\jsobj{}}}} \to
          E_f,H_f
        }
      }
      {
        E, H, \seqq{(E',H')}{\jsfieldass{\jsid{x}}{f}{\jsobj{}}; \jsif{\jscom{false}}{\jsass{y}{\jsfield{\jsid{x}}{f}}}{\jsass{y}{\jsobj{}}}} \to
        E_f,H_f
      }
    }{
      E, H, \jsexpr{\jsass{x}{\jsobj{}}; \jsfieldass{\jsid{x}}{f}{\jsobj{}}; \jsif{\jscom{false}}{\jsass{y}{\jsfield{\jsid{x}}{f}}}{\jsass{y}{\jsobj{}}}}
      \to E_f,H_f
    }
  \end{mathpar}
  \caption{Pretty-big-step derivation}
  \label{fig:pretty_example}
\end{figure}

\section{Annotated Semantics}
\label{sec:instrumentation}

\subsection{Execution traces}
\label{sec:traces}

We want to track how data created at one point in the execution flows into
locations (variables or object fields) at another, later point in the execution
of the program.  To this end, we need a mechanism for talking about ``points of
time'' in a program execution. This information is implicit in the semantic
derivation tree corresponding to the execution. To make it explicit, we
instrument the semantics to produce a (linear) trace of the inference rules used
in the derivation, and use it to refer to particular points in the
execution. Notice that this instrumentation adds no information to the
instrumented trace. We will use these traces later in this section to define
direct flows.

More precisely, we add partial traces, $\tau \in \Trace$, to both sides of the
reduction rules. These traces are lists of rule names decorated with ``$i$'' or
``$o$'', $i$ meaning that we entered in the context of the rule and $o$ that we
finished executing it.  Given a rule whose name is \textsc{R} and an initial
trace $\tau$, we put $\tau; \irule{R}$ on the left of the rule. If the rule is
an axiom (such as \textsc{Cst}), we then put $\tau; \irule{R}; \orule{R}$ on the
right-hand side of the rule. Otherwise we recursively call this annotation
algorithm on the first premise, then thread its result on the second premise if
there is one. The final trace on the right is the one returned by the last
premise, with $\orule{R}$ appended at the end. As each rule appends its name in
the annotation on both sides, they can be uniquely identified. An example of a
derivation with explicitly written traces is given in
Figure~\ref{fig:simpleannotatingexample}.

\begin{figure}
\centering
\begin{subfigure}{\textwidth}
  \centering
    \begin{mathpar}
    \inferrule*[Right=Seq]{
      \inferrule*[leftskip=7em,Left=Asg]{
        \inferrule*[Left=Cst]{ }
        {E, H, \jsexpr{\jscom{true}} \to E, H, \jscom{true}} \\
        \inferrule*[Right=Asg1]
        {E' = E[\jsid{x} \mapsto \jscom{true}]}
        {E,H,\jsasss{x}{(E,H,\jscom{true})} \to E',H}
      }{E, H, \jsexpr{\jsid{x}~\jscom{=}~\jscom{true}} \to E', H} \\
      \inferrule*[Right=Seq1,leftskip=10em,rightskip=10em,vdots=5em]{
        \inferrule*[Right=Asg]{
          \inferrule*[Left=Var]{E'[\jsid{x}] = \jscom{true}}
          {E', H, \jsexpr{x} \to E', H, \jscom{true}}\\
          \inferrule*[Right=Asg1]
          {E'' = E'[\jsid{y} \mapsto \jscom{true}]}
          {E',H,\jsasss{x}{E',H,\jscom{true}} \to E'',H}
        }
        {E', H, \jsexpr{\jsid{y}~\jscom{=}~\jsid{x}} \to E'', H}
      }
      {E, H, \jsexpr{(E',H) \nsem \jsid{y}~\jscom{=}~\jsid{x}} \to E'', H}
    }{E, H, \jsexpr{\jsid{x}~\jscom{=}~\jscom{true}; \jsid{y}~\jscom{=}~\jsid{x}} \to E'', H} 
  \end{mathpar}
    \caption{Unannotated Derivation}
    \label{fig:simpleannotatingexample:withoutannotations}
\end{subfigure}
\begin{subfigure}{\textwidth}
  \centering
   \begin{mathpar}
     \tau_1 = \trace{\irule{Seq}} \and
     \tau_2 = \trace{\irule{Seq}; \irule{Asg}} \and
     \tau_3 = \tau_2 \concat \trace{\irule{Cst}} \and
     \tau_4 = \tau_3 \concat \trace{\orule{Cst}} \and
     \tau_5 = \tau_4 \concat \trace{\irule{Asg1}} \and
     \tau_6 = \tau_5 \concat \trace{\orule{Asg1}} \and
     \tau_7 = \tau_6 \concat \trace{\orule{Asg}} \and
     \tau_8 = \tau_7 \concat \trace{\irule{Seq1}} \and
     \tau_9 = \tau_8 \concat \trace{\irule{Asg}} \and
     \tau_{10} = \tau_9 \concat \trace{\irule{Var}} \and
     \tau_{11} = \tau_{10} \concat \trace{\orule{Var}} \and
     \tau_{12} = \tau_{11} \concat \trace{\irule{Asg1}} \and
     \tau_{13} = \tau_{12} \concat \trace{\orule{Asg1}} \and
     \tau_{14} = \tau_{13} \concat \trace{\orule{Asg}} \and
     \tau_{15} = \tau_{14} \concat \trace{\orule{Seq1}} \and
     \tau_{16} = \tau_{15} \concat \trace{\orule{Seq}} \and
    \inferrule*[Right=Seq]{
      \inferrule*[leftskip=9em,Left=Asg]{
        \inferrule*[Left=Cst]{ }
        {\tau_3, E, H,  \jsexpr{\jscom{true}} \to \tau_4, E, H,  \jscom{true}} \\
        \inferrule*[Right=Asg1]
        {E' = E[\jsid{x} \mapsto \jscom{true}]}
        {\tau_5, E,H,\jsasss{x}{(E,H,\jscom{true})} \to \tau_6, E',H}
      }{\tau_2, E, H,  \jsexpr{\jsid{x}~\jscom{=}~\jscom{true}} \to \tau_7, E', H} \\
      \inferrule*[Right=Seq1,leftskip=19em,rightskip=8em,vdots=5em]{
        \inferrule*[Right=Asg]{
          \inferrule*[Left=Var]{E'[\jsid{x}] = \jscom{true}}
          {\tau_{10}, E', H,  \jsexpr{x} \to \tau_{11}, E', H,  \jscom{true}}\\
          \inferrule*[Right=Asg1]
          {E'' = E'[\jsid{y} \mapsto \jscom{true}]}
          {\tau_{12}, E',H, \jsasss{x}{E',H,\jscom{true}} \to \tau_{13}, E'',H}
        }
        {\tau_9, E', H,  \jsexpr{\jsid{y}~\jscom{=}~\jsid{x}} \to \tau_{14}, E'', H}
      }
      {\tau_8, E, H,  \jsexpr{(E',H) \nsem \jsid{y}~\jscom{=}~\jsid{x}} \to \tau_{15}, E'', H}
    }{\tau_1, E, H,  \jsexpr{\jsid{x}~\jscom{=}~\jscom{true};
        \jsid{y}~\jscom{=}~\jsid{x}} \to \tau_{16}, E'', H}
  \end{mathpar}
    \caption{Derivation Annotated With Traces}
    \label{fig:simpleannotatingexample:withannotations}
\end{subfigure}
  \caption{Annotating A Simple Derivation}
  \label{fig:simpleannotatingexample}
\end{figure}

Since traces uniquely identify places in a derivation, we use them
from now on to refer to states or further instrumentation in the
derivation. More precisely, if $\tau$ is a trace in a given
derivation, we write $E_\tau$ and $H_\tau$ for the environment and
heap at that place.

\subsection{A General Scheme to Define Annotations}
\label{sec:scheme}

In principle, the annotation process takes as argument a full derivation tree
and returns an annotated tree. However, every annotation process we define in
the following, as well as the one deriving traces, can be described by an
iterative process that takes as arguments previous annotations and the
parameters of the rule applied, and returns an annotated rule.

More precisely, our iterative process is based on steps of five kinds:
\audn{init} steps (for every rule), that create the annotation on the left of
the rules based on the annotation passed in argument, \audn{axiom} steps (for
axioms), that transform the annotations on the left of axiom rules into
annotations on the right of the rule, \audn{up} steps (for rules with inductive
premises), that propagate an annotation on the left of a rule to its first
premise, \audn{down} steps (for rules with inductive premises), that propagate
an annotation on the right of the last premise to the right of the rule, and
\audn{next} steps (for rules with two inductive premises), that propagate the
annotations from the left of the current rule and from the right of the first
premise into the left of the second premise. As we are using a pretty-big-step
semantics, there are at most two inductive premises above each rule, thus these
steps are sufficient.

  \newcommand\annotPoint[2]{\tikz[baseline] \node (#1) {$#2$};}
  \newcommand\trAnnot[4][left]{%
    \draw[draw, very thick, ->] (#2) edge[bend #1] node {$#4$} (#3)
  }
  \begin{mathpar}
	\inferrule*{
		\inferrule*{ }{
			\color{blue}\annotPoint{a1}{a_1}\to\annotPoint{a2}{a_2}} \\
		\inferrule*{
			\inferrule*{ }{\color{darkGreen}\annotPoint{a4}{a_4}\to\annotPoint{a5}{a_5}}
		}{\color{red}\annotPoint{a3}{a_3}\to\annotPoint{a6}{a_6}}
	}{\annotPoint{a0}{a_0}\to\annotPoint{a7}{a_7}}
  \end{mathpar}
	\begin{tikzpicture}[overlay, auto]
		\trAnnot[left = 70, color = blue]{a1}{a2}{\color{blue}axiom};
		\trAnnot[left = 70, color = darkGreen]{a4}{a5}{\color{darkGreen}axiom};
		\trAnnot{a0}{a1}{up};
		\trAnnot[left = 70, color = red]{a3}{a4}{\color{red}up};
		\trAnnot[left = 70, color = red]{a5}{a6}{\color{red}down};
		\trAnnot{a6}{a7}{down};
		\trAnnot[right = 5]{a0}{a3}{};
		\trAnnot[right]{a2}{a3}{next};
	\end{tikzpicture}

  This generic approach allows to compose complex annotations, building upon
  previously defined ones.  This general scheme is summed up in the previous
  picture, where each $a_i$ represents an annotation. As \audn{init} steps are
  defined for every rule, they are not depicted. The colors show which steps
  are associated to which rules: $a_0$ is created by the \audn{init} step of the
  bottom rule (black), then it is transformed and control is passed to the
  left axiom rule (black \audn{up}), which applies its \audn{init} step to
  create the blue $a_1$. The blue \audn{axiom} step creates $a_2$, and control
  returns to the bottom rule, where the black \audn{next} step combines $a_1$ and
  $a_0$ to pass it to the red rule. Annotations are propagated in the right
  premise, and ultimately control comes back to the black rule which pulls the
  $a_6$ annotation from the red rule and creates its $a_7$ annotation.

Note that the types of the annotations on the left and the right of the rules do
not have to be the same, as long as every left-hand side annotation has the same
type, and the same for right-hand side annotations.

As an example, we define the \audn{init}, \audn{axiom}, \audn{up}, \audn{down},
and \audn{next} steps corresponding to the addition of partial traces
(see Figure~\ref{fig:annotatingrulesvar:traces} and
\ref{fig:annotatingrulesassign:traces}).
\begin{itemize}
\item The \audn{init} step is defined for every rule and can be seen in both
  figures. Assuming the rule name is \textsc{Name}, then this step appends
  $\irule{Name}$ to the previous trace $\tau$, written $\tau \concat
  \trace{\irule{Name}}$.
\item The \audn{axiom} step only needs to be defined for rules with no inductive
  premise, namely
  \textsc{Skip}, \textsc{WhileFalse}, \textsc{Asg1}, \textsc{FldAsg2}, \textsc{Del1}, \textsc{Cst},
  \textsc{Var}, \textsc{Bin2}, \textsc{Obj}, \textsc{Fld1}, and \textsc{Abort}. It
  adds the current rule name to its argument $\tau$, decorated with
  ``$o$'', as illustrated in Figure~\ref{fig:annotatingrulesvar:traces}: $\tau
  \concat \trace{\orule{Name}}$.
\item The \audn{up} step is only defined on rules that have an inductive
  premise, and it simply propagates the annotation to the first premise. It is
  illustrated in Figure~\ref{fig:annotatingrulesassign:traces}.
\item The \audn{down} step takes the right trace from the last premise of the
  rule above and adds the current rule name to the right annotation decorated
  with ``$o$'' (Figure~\ref{fig:annotatingrulesassign:traces}).
\item The \audn{next} step takes two arguments: the right trace of the first
  premise $\tau_1$ and the left trace $\tau_0$ of the current rule. It ignores
  $\tau_0$ and propagates $\tau_1$ (Figure~\ref{fig:annotatingrulesassign:traces}).
\end{itemize}

\subsection{Dependency Relation}
\label{sec:dependencyrelation}

We are interested in deriving the dependency analysis underlying
tainting analyses for checking that secret values do not flow into
other values that are rendered public. To this end, we consider
\emph{direct flows} from \emph{sources} to \emph{stores}.  We
need a mechanism for describing when data was created and when a flow
happened, so we annotate locations in the
heap with the time when they were allocated. By ``time'' we here mean
the point of time in an execution, represented by a trace $\tau$ of the
derivation.
We write $\ALoc = \Loc \times \Trace$ for the set of \emph{annotated
  locations}.  Similarly, we annotate variables and fields with
the point in time that they were last assigned to.
%
%
When describing a flow, we talk about \emph{sources} and
\emph{stores}. 
Sources are of three kinds:
\begin{itemize}
\item an annotated location, 
\item a variable annotated with
its last modification time, 
\item or a pair of annotated location and field
further annotated with their last modification time. 
\end{itemize}
Stores are either a variable or a pair of an annotated location and a
field, further annotated with their last modification time.  Formally, we define the following
dependency relation
\[
\flowitem \in \Dep = \powerset[\Source \times \Stor]
\]
where $\Stor = \left(\Var \times \Trace\right) + \left(\ALoc \times
  \Field \times \Trace\right)$ and $\Source = \ALoc + \Stor$.

For instance, we write $\jsid[\tau_1]{y} \flowitem \jsid[\tau_2]{x}$
to indicate that the content that was put in the variable $\jsid{y}$
at time $\tau_1$ has been used to compute the value stored in the
variable $\jsid{x}$ at time $\tau_2$. Similarly, we write $\aloc{l}{\tau_2}
\flowitem \locfield[\tau_3]{\aloc{l'}{\tau_1}}{f}$ to indicate that the
object allocated at location $l$ at time $\tau_2$ flows at time $\tau_3$ into
field $\jsid{f}$ of location $l'$ that was allocated at time $\tau_1$.

\begin{figure}[tb]
	\centering
	\begin{subfigure}{0.4\textwidth}
		\centering
		\begin{mathpar}
			\inferrule*[Left=Var]{
			  E[\jsexpr{\jsid{x}}] = v
			}{
			  E, H, \jsexpr{\jsid{x}} \to E, H, v
			}
		\end{mathpar}
		\caption{Basic Rule}
		\label{fig:annotatingrulesvar:basic}
	\end{subfigure}
	\quad
	\begin{subfigure}{0.55\textwidth}
		\centering
		\begin{mathpar}
			\inferrule*[Right=Var]{
			  \grey{E[\jsexpr{\jsid{x}}] = v} \\
			  \tau' = \tau \concat \trace{\irule{Var};\orule{Var}}
			}{
			  \grey{E, H, } \tau \concat \trace{\irule{Var}} \grey{, \jsexpr{\jsid{x}}} \to \grey{E, H, } \tau' \grey{, v}
			}
		\end{mathpar}
		\caption{Adding Partial Traces}
		\label{fig:annotatingrulesvar:traces}
	\end{subfigure}
	\quad
	\begin{subfigure}{0.4\textwidth}
		\centering
		\begin{mathpar}
			\inferrule*[Left=Var]{
			  \grey{E[\jsexpr{\jsid{x}}] = v} \\
			  \grey{\tau' = \tau \concat \trace{\irule{Var};\orule{Var}}}
			}{
			  \grey{E, H, \tau \concat{\irule{Var}}, } M \grey{, \jsexpr{\jsid{x}}} \to \grey{E, H, \tau',} M \grey{, v}
			}
		\end{mathpar}
		\caption{Adding Last-Modified Place}
		\label{fig:annotatingrulesvar:lc}
	\end{subfigure}
	\quad
	\begin{subfigure}{0.55\textwidth}
		\centering
		\begin{mathpar}
			\inferrule*[Right=Var]{
			  \grey{E[\jsexpr{\jsid{x}}] = v} \\
			  \grey{\tau' = \tau \concat \trace{\irule{Var}; \orule{Var}}} \\
			  M[\jsid{x}] = \tau_0
			}{
			  \grey{E, H, \tau \concat{\irule{Var}}, M, \jsexpr{\jsid{x}}} \to \grey{E, H, \tau', M,} \mset{\jsid{x}^{\tau_0}} \grey{, v}
			}
		\end{mathpar}
		\caption{Adding Dependencies}
		\label{fig:annotatingrulesvar:dep}
	\end{subfigure}
	\caption{Instrumentation Steps for \textsc{Var}}
	\label{fig:annotatingrulesvar}
\end{figure}

\begin{figure}[tb]
	\centering
	\begin{subfigure}{\textwidth}
		\centering
		\begin{mathpar}
			\inferrule*[Right=Asg]{
			  \inferrule*[Left=Rule]{\vdots}{S, e \to r} \\
        \inferrule*[Right=Asg1]{\vdots}{S, \jsasss{x}{r} \to r'}
			}{
			  S, \jsass{x}{e} \to r'
			}
		\end{mathpar}
		\caption{Basic Rule}
		\label{fig:annotatingrulesassign:basic}
	\end{subfigure}
	\quad
	\begin{subfigure}{\textwidth}
		\centering
		\begin{mathpar}
      \tau_0 = \tau \concat{\trace{\irule{Asg}}} \and
      \tau_1 = \tau_0 \concat \trace{\irule{Rule}} \and
      \tau_3 = \tau_2 \concat \trace{\irule{Asg1}} \and
      \tau_5 = \tau_4 \concat{\trace{\orule{Asg}}} \and
			\inferrule*[Right=Asg]{
			  \inferrule*[Left=Rule]{\vdots}{\tau_1 \grey{,S,e} \to \tau_2 \grey{,r}} \\
        \inferrule*[Right=Asg1]{\vdots}{\tau_3
          \grey{, S, \jsasss{x}{r}} \to \tau_4 \grey{,r'}}
			}{
			  \tau_0 \grey{, \jsass{x}{e},} \to
         \tau_5 \grey{, r'}
			}
		\end{mathpar}
		\caption{Adding Partial Traces}
		\label{fig:annotatingrulesassign:traces}
	\end{subfigure}
	\quad
	\begin{subfigure}{\textwidth}
		\centering
		\begin{mathpar}
			\inferrule*[Right=Asg]{
			  \inferrule*[Left=Rule]{\vdots}{\grey{\tau_1,} M \grey{,S,e} \to
          \grey{\tau_2,} M' \grey{,r}} \\
        \inferrule*[Right=Asg1]{\vdots}{\grey{\tau_3, } M'
          \grey{, S, \jsasss{x}{r}} \to \grey{\tau_4, } M'' \grey{,r'}}
			}{
			  \grey{\tau_0,} M \grey{, \jsass{x}{e},} \to
         \grey{\tau_5,} M'' \grey{, r'}
			}
		\end{mathpar}
		\caption{Adding Last-Modified Place}
		\label{fig:annotatingrulesassign:lc}
	\end{subfigure}
	\quad
	\begin{subfigure}{\textwidth}
		\centering
		\begin{mathpar}
			\inferrule*[Right=Asg]{
        \grey{\tau_1, M,  \emptyset, } \Delta \grey{,S,e} \to \grey{\tau_2, M',
          d ,} \Delta \grey{,r} \\
        \grey{\tau_3,  M', d, } \Delta \grey{, S, \jsasss{x}{r}} \to \grey{\tau_4,  M'',
        \emptyset, } \Delta' \grey{,r'}
			}{
			  \grey{\tau_0, M,  \emptyset, } \Delta \grey{,S, \jsass{x}{e}} \to
         \grey{\tau_5, M'', \emptyset, } \Delta' \grey{, r'}
			}
		\end{mathpar}
		\caption{Adding Dependencies}
		\label{fig:annotatingrulesassign:dep}
	\end{subfigure}
	\caption{Instrumentation Steps for \textsc{Asg}}
	\label{fig:annotatingrulesassign}
\end{figure}

\subsection{Direct Flows}
\label{sec:instrumentation:flows}

We now detail how to compose additional annotations to define our
direct flow property $\flowitem$. As flows are a global property of the
derivation, we use a series of annotations to propagate local information until
we can locally define direct flows.

We first collect in the derivation the traces where locations are
created and where variables or object fields are assigned. To this
end, we define a new annotation $M$ of type $\left(\Loc + \Var + \ALoc
\times \Field\right) \to \Trace$. After this instrumentation step, reductions
are of the form $\tau, M_\tau, S_\tau, t \to \tau', M_{\tau'}, r$.

The three rules that modify $M$ are \textsc{Obj}, \textsc{Asg1}, and
\textsc{FldAsg2}. We describe them in Figure \ref{fig:instrumentm}. The other
rules simply propagate $M$. For the purpose of our analysis, we do not consider
the deletion of a field as its modification. More precise analyses, in
particular ones that also track indirect flows, would need to record such
events.

\begin{figure}[tb]
  \centering
  \begin{mathpar}
    \inferrule*[Right=Obj]{
      \grey{H[l] = \bot} \\
      \grey{H' = H[l \mapsto \jsobj{}]} \\
      M' = M[l \mapsto \tau']
    }{
      \grey{\tau,} M \grey{, E, H, \jsexpr{\jsobj{}}} \to \grey{\tau',} M' \grey{, E, H', l}
    }
    \and
    \inferrule*[right=FldAsg2]{
      \grey{H[l] = o}\\
      \grey{o' = o\left[\jsid{f} \mapsto v\right]}\\
      \grey{H' = H'\left[l \mapsto o'\right] }\\
      M' = M[(l, M[l], f) \mapsto \tau']
    }{
      \grey{\tau,} M \grey{, S, \jsfieldassss{l}{f}{(E,H,v)}} \to \grey{\tau',}
      M' \grey{, E, H'}
    }
    \and
    \inferrule*[Right=Asg1]{
      \grey{E' = E[\jsid{x} \mapsto v] }\\
      M' = M[\jsid{x} \mapsto \tau']
    }{
      \grey{\tau,} M \grey{, S, \jsasss{x}{(E,H,v)}} \to \grey{\tau',} M' \grey{, E', H}
    }
  \end{mathpar}
  \caption{Adding Modified and Created Information}
  \label{fig:instrumentm}
\end{figure}

The added instrumentation uses traces to track the moments when
locations are created, and when fields and variables are assigned.
For field assignment, the rule \textsc{FldAsg2} relies on the fact
that the location of the object assigned has already been created
to obtain the annotated location:  we have the invariant that if
$H[l]$ is defined, then $M[l]$ is defined.

We can now continue our instrumentation by adding \emph{dependencies} $d \in
\powerset[\Var]$. The instrumented reduction is now $\tau, M_\tau, d_\tau,
S_\tau, t \to \tau', M_{\tau'}, d_{\tau'}, r$. Its rules are described in
Figure~\ref{fig:annotationexpressionsdependencies}. The rules not given only
propagate the dependencies. The intuition behind these rules is that expressions
generate potential dependencies that are thrown away when they don't result in
direct flow (for instance when computing the condition of a \textsc{If}
statement).  The important rules are \textsc{Var}, where the result depends on
the last time the variable was modified, \textsc{Obj}, which records the
dependency on the creation of the object, and \textsc{Fld1}, whose result
depends on the last time the field was assigned. The \textsc{Asg} and
\textsc{FldAsg1} rules make sure these dependencies are transmitted to the
inductive call to the rule that will proceed with the assignment for the next
series of annotations.

\begin{figure}
	\centering
	\begin{mathpar}
	  \inferrule*[Right=Var]{
	  	\grey{E[\jsid{x}] = v} \\
	  	M[\jsid{x}] = \tau_0
	  }{
	  	\grey{\tau, M,} d \grey{, E, H, \jsexpr{\jsid{x}}} \to \grey{\tau', M,} d
      \cup \mset{\jsid[\tau_0]{x}} \grey{, E, H, v}
	  }
    \and
	  \inferrule*[Right=Obj]{
		\grey{H[l] = \bot} \\
		\grey{H' = H[l \mapsto \jsobj{}]} \\
    \grey{M' = M[l \mapsto \tau']}
	  }
	  {
		\grey{\tau, M,} d \grey{, E, H, \jsexpr{\jsobj{}}} \to \grey{\tau', M,} d
    \cup \mset{\aloc{l}{\tau'}} \grey{, E, H', l}
	  } \and
    \inferrule*[right=Fld1]{\grey{H'[l] = o} \\ \grey{o[\jsid{f}] = v} \\ M[(l, M[l], f)] = \tau_0}
    {\grey{\tau, M, } d \grey{, E, H, \jsfieldd{(E',H',l)}{f}} \to \grey{\tau',
        M, } d \cup \mset{(l, M[l], f)^{\tau_0}} \grey{, E', H', v}}
    \and
    \inferrule*[right=If]{
      \grey{\tau_1, M,} \emptyset \grey{, S, \jsexpr{e}} \to \grey{\tau_2, M',}
      d \grey{, r} \\
      \grey{\tau_3, M',} \emptyset \grey{, S, \jsiff{r}{s_1}{s_2}} \to
      \grey{\tau_4, M'',} \emptyset \grey{, r'}
    }{
      \grey{\tau_0, M,} \emptyset \grey{, S, \jsif{e}{s_1}{s_2}} \to
      \grey{\tau_5, M'',} \emptyset \grey{, r'}
    } \and
    \inferrule*[right=While]{
      \grey{\tau_1, M, } \emptyset \grey{, S, e} \to \grey{\tau_2, M',} d \grey{, r}\\
      \grey{\tau_3, M', } \emptyset \grey{, S, \jswhilee{r}{x}{s}} \to
      \grey{tau_4, M'',} \emptyset \grey{, r}'
    }{
      \grey{\tau_0, M, } \emptyset \grey{,S, \jswhile{e}{s}} \to \grey{\tau_5,
        M'', } \emptyset \grey{, r'}
    } \and
			\inferrule*[Right=Asg]{
        \grey{\tau_1, M, } \emptyset \grey{,S,e} \to \grey{\tau_2, M', } d \grey{,r} \\
        \grey{\tau_3,  M',} d \grey{, S, \jsasss{x}{r}} \to \grey{\tau_4,  M'',}
        \emptyset \grey{,r'}
			}{
			  \grey{\tau_0, M, } \emptyset \grey{,S, \jsass{x}{e}} \to
         \grey{\tau_5, M'',} \emptyset \grey{, r'}
			}
      \and
      \inferrule*[right=Asg1]{
      \grey{E' = E[\jsid{x} \mapsto v]} \\
      \grey{M' = M[\jsid{x} \mapsto \tau']}
    }{
      \grey{\tau, M,} d \grey{, S, \jsasss{x}{(E,H,v)}} \to
      \grey{\tau', M',} \emptyset \grey{, E', H}
    }
    \and
    \inferrule*[Right=FldAsg]{
        \grey{\tau_1, M, } \emptyset \grey{,S,e_1} \to \grey{\tau_2, M', } d \grey{,r} \\
        \grey{\tau_3,  M',} \emptyset \grey{, S, \jsfieldasss{r}{f}{e_2}} \to \grey{\tau_4,  M'',}
        \emptyset \grey{,r'}
			}{
			  \grey{\tau_0, M, } \emptyset \grey{,S, \jsfieldass{e_1}{f}{e_2}} \to
         \grey{\tau_5, M'',} \emptyset \grey{, r'}
			}
    \and
			\inferrule*[Right=FldAsg1]{
        \grey{\tau_1, M, } \emptyset \grey{,S',e} \to \grey{\tau_2, M', } d \grey{,r} \\
        \grey{\tau_3,  M',} d \grey{, S', \jsfieldassss{l}{f}{r}} \to \grey{\tau_4,  M'',}
        \emptyset \grey{,r'}
			}{
			  \grey{\tau_0, M, } \emptyset \grey{,S, \jsfieldasss{(S',x)}{f}{e}} \to
         \grey{\tau_5, M'',} \emptyset \grey{, r'}
			}
      \and
    \inferrule*[right=FldAsg2]{
      \grey{H[l] = o}\\
      \grey{o' = o\left[\jsid{f} \mapsto v\right]}\\
      \grey{H' = H'\left[l \mapsto o'\right]} \\
      \grey{M' = M[(l, M[l], f) \mapsto \tau']}
    }{
      \grey{\tau, M,} d \grey{, S, \jsfieldassss{l}{f}{(E,H,v)}} \to
      \grey{\tau', M',} \emptyset \grey{, E, H'}
    }
    \and
			\inferrule*[Right=Delete]{
        \grey{\tau_1, M, } \emptyset \grey{,S,e} \to \grey{\tau_2, M', } d \grey{,r} \\
        \grey{\tau_3,  M',} \emptyset \grey{, S, \jsdeletee{r}{f}} \to \grey{\tau_4,  M'',}
        \emptyset \grey{,r'}
			}{
			  \grey{\tau_0, M, } \emptyset \grey{,S, \jsdelete{e}{f}} \to
         \grey{\tau_5, M'',} \emptyset \grey{, r'}
			}
	\end{mathpar}
	\caption{Rules for Dependencies Annotations}
	\label{fig:annotationexpressionsdependencies}
\end{figure}

Finally, we build upon this last instrumentation to define flows. The final
instrumented derivation is of the form:
$\tau, M_\tau, d_\tau, \Delta_\tau, S_\tau, s \to \tau', M_{\tau'}, d_{\tau'}, \Delta_{\tau'}, r_{\tau'}$,
where $\mset{\Delta_\tau, \Delta_{\tau'}} \subseteq \Dep$ are sets of flows defining the
$\flowitem$ relation (see Section~\ref{sec:dependencyrelation}). The
two important rules are \textsc{Asg1} and \textsc{FldAsg2}, which
modify respectively a variable and a field, and for which the flow
needs to be added. All the other rules just propagate those new
constructions. The two modified rules are given in Figure \ref{fig:annotstmtdep}. 

\begin{figure}[tb]
  \centering
  \begin{mathpar}
    \inferrule*[right=Asg1]{
      \grey{E' = E[\jsid{x} \mapsto v]} \\
      \grey{M' = M[\jsid{x} \mapsto \tau']}
    }{
      \grey{\tau, M, d, } \Delta \grey{, S, \jsasss{x}{(E,H,v)}} \to
      \grey{\tau', M', \emptyset, } \mset{\delta \flowitem \jsid{x}^{\tau'} | \delta \in
        d} \cup \Delta \grey{, E', H}
    }
    \and
    \inferrule*[right=FldAsg2]{
      \grey{H[l] = o}\\
      \grey{o' = o\left[\jsid{f} \mapsto v\right]}\\
      \grey{H' = H'\left[l \mapsto o'\right]} \\
      \grey{M' = M[(l, M[l], f) \mapsto \tau']}
    }{
      \grey{\tau, M, d, } \Delta \grey{, S, \jsfieldassss{l}{f}{(E,H,v)}} \to
      \grey{\tau', M', \emptyset, } \mset{\delta \flowitem (l, M[l], f)^{\tau'}
        | \delta \in d} \cup \Delta \grey{, E, H'}
    }
\end{mathpar}
  \caption{Rules for Annotating Dependencies of Statements}
  \label{fig:annotstmtdep}
\end{figure}

\subsection{Correctness Properties of the Annotations}
\label{sec:annotationscorrect}

The instrumentation of the semantics does not add information to the
reduction but only makes existing information explicit. The
correctness of the instrumentation can therefore be expressed as a
series of coherence properties between the instrumented semantics. 

\newcommand{\ppsep}{/}
\newcommand{\ppn}{\textrm{PP}}
\newcommand{\ppt}{\mbox{\tiny{\ppn}}}
\newcommand{\pppt}{\mbox{\tiny{$\ppn'$}}}
\newcommand{\pp}[1]{\textsc{#1}}
\newcommand{\ppe}[1]{\ppsep\textsc{#1}}
\newcommand{\ppb}[1]{\textsc{#1}\ppsep}
\newcommand{\ppet}[1]{\mbox{\tiny{\ppsep\textsc{#1}}}}
\newcommand{\ppf}[2][]{\Pi(\ifthenelse{\equal{#1}{}}{\cdot}{\ppn #1},#2)}

We first describe the relation between program points and traces in
the derivations, and how to add program points to programs. The transformation $\Pi$
described below takes a program point and a term, and annotates each
sub-term with program points before and after the sub-term. The program point
before is a \emph{context}, a list of atoms indicating where to find
the term in the
initial program. The program point after is a context followed by the
name of the syntactic construct corresponding to the sub-term. For instance, the
program point \pp{Seq2}\ppe{Seq2}\ppe{IfE} refers to the point before
\jsexpr{\jscom{false}} in the term $\jsexpr{\jsass{x}{\jsobj{}};
  \jsfieldass{\jsid{x}}{f}{\jsobj{}};
  \jsif{\jscom{false}}{\jsass{y}{\jsfield{\jsid{x}}{f}}}{\jsass{y}{\jsobj{}}}}$;
and \pp{Seq2}\ppe{Seq2}\ppe{IfE}\ppe{Cst} to the point after. The
notion of ``before'' and ``after'' a program point is standard in data
flow analysis, but is here given a semantics-based definition.  

We write $\cdot$ for the empty program context and assume that $\cdot\ppe{Name}$
is equal to $\pp{Name}$.

\begin{align*}
    \ppf{\jsexpr{\jscom{skip}}} &= \jsexpr{\ppn, \jscom{skip}, \ppn\ppe{Skip}}\\
    \ppf{\jsexpr{s_1; s_2}} &= \jsexpr{\ppn, \ppf[\ppe{Seq1}]{s_1};\ppf[\ppe{Seq2}]{s_2},
      \ppn\ppe{Seq}}\\
    \ppf{\jsif{e}{s_1}{s_2}} &= \jsexpr{\ppn, \jsif{\ppf[\ppe{IfE}]{e}}{\ppf[\ppe{If1}]{s_1}}{\ppf[\ppe{If2}]{s_2}},
      \ppn\ppe{If}}\\
\ppf{\jswhile{e}{s}} &= \jsexpr{\ppn, \jswhile{\ppf[\ppe{WhileE}]{e}}{\ppf[\ppe{WhileS}]{s}}, \ppn\ppe{While}}\\
\ppf{\jsass{x}{e}} &= \jsexpr{\ppn, \jsass{x}{\ppf[\ppe{AsgE}]{e}}, \ppn\ppe{Asg}}\\
\ppf{\jsfieldass{e_1}{f}{e_2}} &= \jsexpr{\ppn, \jsfieldass{\ppf[\ppe{FldAsg1}]{e_1}}{f}{\ppf[\ppe{FldAsg2}]{e_2}}, \ppn\ppe{FldAsg}}\\
\ppf{\jsdelete{e}{f}} &= \jsexpr{\ppn, \jsdelete{\ppf[\ppe{DelE}]{e}}{f}, \ppn\ppe{Del}}\\
\ppf{\jsexpr{c}} &= \jsexpr{\ppn, \jsexpr{c}, \ppn\ppe{Cst}}\\
\ppf{\jsexpr{\jsid{x}}} &= \jsexpr{\ppn, \jsexpr{\jsid{x}}, \ppn\ppe{Var}}\\
\ppf{\jsop{e_1}{e_2}} &= \jsexpr{\ppn, \jsop{\ppf[\ppe{Bin1}]{e_1}}{\ppf[\ppe{Bin2}]{e_2}}, \ppn\ppe{Bin}}\\
\ppf{\jsexpr{\jsobj{}}} &= \jsexpr{\ppn, \jsexpr{\jsobj{}}, \ppn\ppe{Obj}}\\
\ppf{\jsfield{e}{f}} &= \jsexpr{\ppn, \jsfield{\ppf[\ppe{FldE}]{e}}{f}, \ppn\ppe{Fld}}
\end{align*}

\newcommand{\ttoppn}{\mathcal{T}}
\newcommand{\ttopp}[3]{\ttoppn(#1,#2,#3)}

We next define a function $\ttoppn$ from traces, list of unmatched trace
points, and partial program points to program points and unmatched traces. In the
following we write $\_$ for any trace atom, as long as it has not been matched
by a previous rule. This function relies on some additional helper functions whose
rules can be found in Figure \ref{fig:tracespphelper} at the end of the paper.

\newcommand{\appi}[2]{\texttt{App}_{i}(\textsc{#1},#2)}
\newcommand{\appo}[2]{\texttt{App}_{o}(\textsc{#1},#2)}
\newcommand{\pusho}[1]{\texttt{Push}_o(\textsc{#1})}
\newcommand{\pushi}[1]{\texttt{Push}_i(\textsc{#1})}

\begin{align*}
  \ttopp{\tau \concat \trace{\orule{Name}}}{[]}{\ppn} &=
  \ttopp{\tau}{\pusho{Name} :: []}{\appo{Name}{\ppn}}\\
  \ttopp{\tau \concat \trace{\orule{Name}}}{\orule{Name} :: l}{\ppn} &=
  \ttopp{\tau}{\orule{Name} :: \orule{Name} :: l}{\ppn}\\
  \ttopp{\tau \concat \trace{\irule{Name}}}{\orule{Name} :: l}{\ppn} &=
  \ttopp{\tau}{l}{\ppn} \\
  \ttopp{\tau \concat \trace{\_}}{\orule{Name} :: l}{\ppn} &=
  \ttopp{\tau}{\orule{Name} :: l}{\ppn}\\
  \ttopp{\tau \concat \trace{\irule{Name}}}{[]}{\ppn} &= \ttopp{\tau}{\pushi{Name}}{\appi{Name}{\ppn}}
\end{align*}

We call ``normal names'' the name of rules for the non-extended terms, \emph{i.e.}
\textsc{Skip}, \textsc{Seq}, \textsc{If}, \textsc{While}, \textsc{Asg},
\textsc{FldAsg}, \textsc{Del}, \textsc{Cst}, \textsc{Var}, \textsc{Bin},
\textsc{Obj}, and \textsc{Fld}. We call ``extended names'' the names of the
other rules.

We now state that program points can correctly be extracted from traces. To
this end, we consider a derivation where the terms contain program points. Note
that only normal terms have exposed program points, of the form $\ppn, t,
\ppn'$. Extended terms also contain program points but have none at toplevel.

\begin{prop}\label{prop:pp}
  Let $t$ be a term and $t' = \ppf{t}$. For any occurrence of a normal
  rule \textsc{Name} of the form $\tau, S, \ppn,t,\ppn' \to \tau',r$ in any
  annotated derivation tree from $t'$, we have $\ppn = \ttopp{\tau}{[]}{\cdot}$,
  $\ppn' = \ttopp{\tau'}{[]}{\cdot}$, and $\ppn' = \ppn\ppe{Name}$.
\end{prop}

We next state correctness properties about the instrumentation of the heap.  We
start by a property concerning the last-modified-place
annotations. This property states that the annotation of a location's
creation point never changes, and that the value of a field has not
changed since the point of 
  modification indicated by the instrumentation component $M$.

\begin{prop}
	For every instrumented derivation tree, and for every rule in this tree
\[
	\tau, M_\tau, E_\tau, H_\tau, t \to \tau', M_{\tau'}, r
  \] where $\st{r} = E_{\tau'}, H_{\tau'}$ and
  $M_{\tau'}\left[\locfield{\aloc{l}{\tau_0}}{f}\right] = \tau_1$, Then we have
  $M_{\tau'}[l] = \tau_0$ and $H_{\tau'}[l][\jsid{f}] =
  H_{\tau_1}[l][\jsid{f}]$.
\end{prop}

The following property links the last-change-place annotation ($M$)
with the dependencies annotation ($\Delta$). Intuitively, it states
that if $\Delta$ says that the value assigned to $x$ at time $\tau_1$
later flew into a variable a time $\tau_2$ then $x$ has not changed
between $\tau_1$ and $\tau_2$. 
\begin{prop}
	For every instrumented derivation tree, and for every rule in this tree
\[
	s, \tau, M_\tau, d_\tau, \Delta_\tau, S_\tau, t \to \tau', M_{\tau'},
  d_{\tau'}, \Delta_{\tau'}, r
\] 
if $\jsid[\tau_1]{x} \flowitem \jsid[\tau_2]{y} \in \Delta_{\tau'}$, then at
time $\tau_2$, the last write to \jsid{x} was at time $\tau_1$, \emph{i.e.}
$M_{\tau_2}[\jsid{x}] = \tau_1$.
\end{prop}


We now state the most important property: if at some point during the execution
of a program the field of an object contains another object, then there is a
chain of direct flows attesting it in the annotation.

More precisely, we write $l_0 \flowitem^*_\Delta l_n.\jsid{f}$ if there are
stores $s_0$ \ldots $s_n$ such that:
\begin{itemize}
\item $s_0 = \aloc{l_0}{\tau_0}$ for some $\tau_0$, $s_n =
  \locfield[\tau'_n]{\aloc{l_n}{\tau_n}}{f}$ for some $\tau_n$ and $\tau'_n$,
  and for every $i$ in $[1..n]$ we have either
  $s_i=\locfield[\tau'_i]{\aloc{l_i}{\tau_i}}{f$_i$}$ for some $l_i$, $\jsid{f}_i$,
  $\tau_i$, and $\tau'_i$ or $s_i = \jsid[\tau'_i]{x$_i$}$ for some $\jsid{x}_i$ and
  $\tau'_i$; and
\item  for every $i$, $s_i \flowitem s_{i + 1} \in \Delta$. 
\end{itemize}

\begin{prop}
	For every instrumented derivation tree, and for every rule in this tree
\[
	 \tau, M_\tau, d_\tau, \Delta_\tau, E_\tau, H_\tau, t \to \tau', M_{\tau'}, d_{\tau'},
   \Delta_{\tau'}, r
\]
 where $\st{r} = E_{\tau'},H_{\tau'}$, we have:
 \begin{itemize}
 \item for every locations $l,l'$ and field $\jsid{f}$ such that
   $H_\tau[l'][\jsid{f}] = l$, then $l \flowitem^*_{\Delta_\tau} l'.\jsid{f}$,
 \item for every locations $l,l'$ and field $\jsid{f}$ such that
   $H_{\tau'}[l'][\jsid{f}] = l$, then $l \flowitem^*_{\Delta_{\tau'}} l'.\jsid{f}$.
 \end{itemize}
\end{prop}

\subsection{Annotated Semantics in Coq}
\label{sec:annotatedsemanticincoq}

In the \coqn development, we distinguish expressions from
statements, and we define the reduction $\to$ as two coq
predicates: \coqinline|red_expr| and
\coqinline|red_stat|. The first predicate has type
\coqinline|environment -> heap -> ext_expr -> out_expr -> Type| 
(and similarly for the statement reduction).  The
construction \coqinline|ext_expr| refers to the extended syntax for expressions
$e_e$.  The inductive type \coqinline|out_expr| is defined as being either the
result of a terminating evaluation, containing a new environment, heap, and
returned value, or an aborted evaluation, containing a new environment and heap.

\begin{coq}
Inductive out_expr :=
  | out_expr_ter : environment -> heap_o -> value -> out_expr
  | out_expr_error : environment -> heap_o -> out_expr.
\end{coq}

To ease the instrumentation, we directly add the annotations in the semantics:
each rule of the semantics takes two additional arguments: the left-hand side
annotation and the right-hand side annotation.  However, there is no restriction
on these annotations, we rely on the correctness properties of Section
\ref{sec:annotationscorrect} to ensure they define the property of interest.

The semantics is thus parametrised by four types, corresponding to the left and
right annotations for expressions and statements.  These types are wrapped in a
\coqn record and used through projections such as \coqinline|annot_e_l| (for
left-hand-side annotations in expressions).

\begin{figure}
\centering
\begin{coq}
Inductive red_expr : environment -> heap_o -> ext_expr -> out_expr -> Type :=

  (* ... *)

  | red_expr_expr_var : annot_e_l Annots -> annot_e_r Annots ->
    forall E H x v,
    getvalue E x v ->
    red_expr E H (ext_expr_expr (expr_var x)) (out_expr E H v)
\end{coq}
\caption{A Semantic Rule as Written in \coqn}
\label{fig:coqredexprvar}
\end{figure}

Figure~\ref{fig:coqredexprvar} shows the rule for variables from this annotated
semantics, where \coqinline|ext_expr_expr| corresponds to the injection of
expressions into extended expressions.  The additional annotation arguments of
type \coqinline|annot_e_l| and \coqinline|annot_e_r| are carried by every rule.
As every rule contains such annotations, it is easy to write a function
\coqinline|extract_anot| taking such a derivation tree and returning the
corresponding annotations.  Every part of the \coqn development that uses the
reduction $\to$ but not the annotations (such as the interpreter) uses trivial
annotations of unit type.

The annotations are then incrementally computed using \coqn functions.  Each of the new annotating passes takes the
result of the previous pass as an argument to add its new annotations.  The
initial annotation is the trivial one, where every annotating types are unit.
The definition of annotations in our \coqn development exactly follows the
scheme presented in Section \ref{sec:scheme}.
This allows to only specify the parts of the analysis that effectively change their annotations, using a pattern matching construction ending with a \coqn's wildcard \coqinline|_| to deal with all the cases that just propagate the annotations.
It has been written in a modular way, which is robust to changes. For
example, a previous version of the annotations only used open rules for the partial traces and not closing one.
As all the following annotating passes treat the traces as an abstract object whose type is parameterized, it was straightforward to update the \coqn development for this change.

\begin{figure}
\centering
\begin{coql}
Section LastModified.
Variable Locations : Annotations.

Definition ModifiedAnnots := annot_s_r Locations.
@\label{fig:lastchangeinCoq:LastModifiedHeaps}@Record LastModifiedHeaps : Type :=
  makeLastModifiedHeaps {
    LCEnvironment : heap var ModifiedAnnots;
    LCHeap : heap loc (heap prop_name ModifiedAnnots)
  }.

Definition LastModified :=
  ConstantAnnotations LastModifiedHeaps. @\label{fig:lastchangeinCoq:constantAnnots}@

Definition LastModifiedAxiom_s (r : LastModifiedHeaps)
  E H t o (R : red_stat Locations E H t o) :=
  let LCE := LCEnvironment r in
  let LCH := LCHeap r in
  let (_, tau) := extract_annot_s R in
  match R with
  | red_stat_ext_stat_assign_1 _ _ _ _ _ _ _ x _ _ => @\label{fig:lastchangeinCoq:caseofassign}@
    let LCE' := write LCE x tau
    in makeLastModifiedHeaps LCE' LCH
  | red_stat_stat_delete _ _ _ _ _ l _ f _ _ _ _ _ _ =>
    let aob := read LCH l
    in let LCH' := write LCH l (write aob f tau)
    in makeLastModifiedHeaps LCE LCH'
  | red_stat_ext_stat_set_2 _ _ _ _ _ _ _ _ l _ f _ _ _ _ _ _ =>
    let aob := read LCH l
    in let LCH' := write LCH l (write aob f tau)
    in makeLastModifiedHeaps LCE LCH'
  | _ => makeLastModifiedHeaps LCE LCH
  end.

@\label{fig:lastchangeinCoq:annotLastModified}@Definition annotLastModified :=
  makeIterativeAnnotations LastModified
    (init_e Transmit) (axiom_e Transmit) (up_e Transmit) (down_e Transmit) (next_e Transmit)
    (up_s_e Transmit) (next_e_s Transmit)
    (init_s Transmit) LastModifiedAxiom_s (up_s Transmit) (down_s Transmit) (next_s Transmit).

End LastModified.
\end{coql}
\caption{\coqn Definitions of the Last-Modified Annotation}
\label{fig:lastchangeinCoq}
\end{figure}

Figure~\ref{fig:lastchangeinCoq} shows the introduction of the last-modified annotation (see Figure~\ref{fig:annotatingrulesvar:lc} and \ref{fig:annotatingrulesassign:lc}).
This annotation is parameterized by another (traces for instance) here called \coqinline|Locations|.
In \coqn, the heap $M$ of Section~\ref{sec:instrumentation:flows} is represented
by the record \coqinline|LastChangeHeaps| defined on
Line~\ref{fig:lastchangeinCoq:LastModifiedHeaps}. Line~\ref{fig:lastchangeinCoq:constantAnnots}
then states it is the left and right annotation types of this annotation.
Next is the pattern matching defining the \audn{axiom} rule for statement, and in particular the case of the assignment Line~\ref{fig:lastchangeinCoq:caseofassign} which, as in Figure~\ref{fig:annotatingrulesassign:lc}, stores the current location $\tau$ in the annotation.
Line~\ref{fig:lastchangeinCoq:annotLastModified} sums up the rules, stating that every rule of this annotation just propagates their arguments, except the \audn{axiom} rule for statements.
As can be seen, the corresponding code is fairly short.

We have also defined an interpreter \coqinline|run_expr : nat -> environment -> heap_o -> expr -> option out|
taking as arguments an integer, an environment, a heap, and an expression and returning an output.
The presence of a \lstinline|while| in \nanoJS allows the existence of
non-terminating executions, whereas every \coqn function must be terminating.
To bypass this mismatch, the interpreters \coqinline|run_expr| and
\coqinline|run_stat| (respectively running over expressions and statements) take
an integer (the first argument of type \coqinline|nat| above), called \emph{fuel}.
At each recursive call, this fuel is decremented, the interpreter giving up and
returning \coqinline|None| once it reaches $0$. We have proven the interpreter
is correct related to the semantics, and we have extracted it as an \caml program
using the \coqn extraction mechanism.

\section{Dependency Analysis}
\label{sec:analysis}

The annotating process makes the property we want to track appear
explicitly in derivation trees. However, the set of properties in question is
still infinite so it might not be possible to compute the instrumented
semantics of a program. The next step is to define an abstraction of
the semantics 
for computing safe approximations of these properties, and to prove
its correctness with respect to the instrumented semantics.

\subsection{Abstract Domains}
\label{sec:analysis:domains}

The analysis is expressed as a reduction relation
operating over abstractions of the concrete semantic domains.
The notion of program point will play a central role, as program
points are used both in the abstraction of points of allocation and
points of modification. 
This analysis thus uses the set $\PP$ of program points, so we assume that the input
program is a result of Function~$\Pi$ defined in
Section~\ref{sec:annotationscorrect}. 
Property~\ref{prop:pp} assures that the added program points are correct with
respect to the associated traces, which are used to name objects, and thus
that this abstraction is sound.
To avoid burdening notations, program points are only shown when needed.

Values were defined to be either basic values or locations. We shall
ignore the basic values and focus on abstracting the unbounded set of heap locations. 
There are several ways of abstracting objects in the heap, but we
shall content ourselves with the standard abstraction in which object
locations are abstracted by the program points corresponding to the
instruction that allocated the object.
Thus
\[
\abs{l} \in \abs\Loc = \powerset[\PP].
\]
The abstraction of values should in addition contain a component for tracking 
variables $\jsid{x}$ on which the value depends. 
Values are thus abstracted by a pair $\abs{v}$ of abstract location
$l$ and of the set $\abs{d}$ of
variables that possibly flowed into this specific value,
annotated with the program points of their definition.
\[
	\abs{v} \in \abs\Val = \abs\Loc \times \powerset[\Var \times \PP]
\]
For objects stored at heap locations we keep trace of the values that
the fields may reference. 
Environments and heaps are then abstracted as follows:
\begin{itemize}
 \item $\abs{E} \in \abs{\Env} = \Var \rightarrow \abs{\Val}$ maps variables to abstract values $\abs{v}$.
 \item $\abs{H} \in \abs{\Heap} = \abs\Loc \rightarrow \Field
   \rightarrow \abs{\Val}$ maps abstract locations to object
   abstractions (that map fields to abstract values). 
\end{itemize}

The two abstract domains inherit a lattice structure in the canonical
way as monotone maps, ordered pointwise. The abstract heaps $\abs{H}$ map
abstract locations $\abs\Loc$ (which are sets of program points)
to abstract object.  As locations are abtracted by sets, each write of a
value $\abs{v}$ in the abstract heap at abstract location $\abs{l}$
implicitely yields a join between $\abs{v}$ and every value associated to
an $\abs{l'} \sqsubseteq \abs{l}$.

We recall the definition of $\Dep$, $\Stor$ and $\Source$:
\begin{mathpar}
\Dep = \powerset[\Source \times \Stor]\and
\ALoc = \Loc \times \Trace\and
\Stor = \left(\Var \times \Trace\right) + \left(\ALoc \times \Field \times \Trace\right)\and
\Source = \ALoc + \Stor
\end{mathpar}
We want to abstract $l^\tau \in \ALoc$ by the program point that allocated
$l^\tau$, and the traces by program points (using $\prec$).
We can thus abstract the relation $\Delta_\tau \in \Dep$ (and the relation
$\flowitem$) by making the natural abstraction $\Delta^\sharp$ of those definitions:
\begin{mathpar}
	\Delta^\sharp \in \abs\Dep = \powerset[\abs\Source \times \abs\Stor]\and
    \abs\ALoc = \PP\\
    \abs\Stor = \left(\Var \times \PP\right) + \left(\PP \times
        \Field \times \PP\right)\and
    \abs\Source = \PP + \abs\Stor
\end{mathpar}
Abstract flows are written using the symbol $\abs\flowitem$.
To avoid confusion, program points $p \in \PP$ interpreted as elements of
$\abs\Source$ (thus representing locations) are written $\objatpp{p}$.

Abstract flows are thus usual flows in which all traces have been
replaced by program points.
We've seen in Section~\ref{sec:traces} that there exists an abstraction
relation $\prec$ between traces and program points such that
$\tau \prec p$ if and only if $p$ corresponds to the trace $\tau$.
This relation can be directly extended to $\Dep$ and $\abs\Dep$:
for instance for each $\jsid[\tau]{x} \in \Var \times \Trace \subset \Stor$
such that $\tau \prec p$, we have $\jsid[\tau]{x} \prec \jsid[p]{x} \in \abs\Stor$.
Similarly, this relation $\prec$ can also be defined over $\Val$ and $\abs\Val$,
$\Env$ and $\abs\Env$, and $\Heap$ and $\abs\Heap$.

\subsection{Abstract Reduction Relation}
\label{sec:analysis:abstractreduction}

We formalize the analysis as an abstract reduction relation $\absto$
for expressions and statements:
\begin{mathpar}
	E^\sharp, H^\sharp, \jsexpr{s} \absto E'^\sharp, H'^\sharp, \Delta^\sharp \and
	E^\sharp, H^\sharp, \jsexpr{e} \absto l^\sharp, d^\sharp
\end{mathpar}
On statements, the analysis returns an abstract environment, an abstract
heap, and a partial dependency relation.  On expressions, it
returns the set of all its possible locations and the set of its
dependencies.
The analysis is correct if, for all
statements, the result of the abstract reduction relation is a correct
abstraction of the instrumented reduction.
More precisely, the analysis is correct if for each statement \jsexpr{s} such that
\[
\tau, M_\tau, d_\tau, \Delta_\tau, E_\tau, H_\tau, \jsexpr{s} \to \tau', M_{\tau'}, d_{\tau'}, \Delta_{\tau'}, E_{\tau'}, H_{\tau'}
\quad \mathrm{and} \quad
E^\sharp, H^\sharp, \jsexpr{s} \absto E'^\sharp, H'^\sharp, \Delta^\sharp
\]
where $M_\tau$ is chosen accordingly to $E_\tau$ and $H_\tau$, $E_\tau \prec \abs{E}$,
and $H_\tau \prec \abs{H}$, we have $\Delta_{\tau'} \prec \Delta^\sharp$.
In other words, the analysis captures at least all the real flows,
defined by the annotations.

Figure~\ref{fig:abstractreduction} shows the rules of this analysis.
To avoid burdening notations, we denote by $d^\sharp \flowitem f$ the
abstract dependency relation $\left\{ (f_d, f) \middle| f_d \in d^\sharp\right\}$.
Following the same scheme, we freely use the notation $l^\sharp\jscom{.}\jsid[p]{f}$
to denote the set $\left\{p_l\jscom{.}\jsid[p]{f}\middle|p_l\in l^\sharp\right)$.
As an example, here is the rule for assignments:
\[
	\inferrule*[right=Asg]{
		E^\sharp, H^\sharp, \jsexpr{e} \absto l^\sharp, d^\sharp
	}{
		E^\sharp, H^\sharp, \jsass[p]{x}{e} \absto E^\sharp\left[\jsid{x} \mapsto \left( l^\sharp, d^\sharp \right)\right], H^\sharp, d^\sharp \abs\flowitem \jsid[p]{x}
	}
\]
This rule expresses that when encountering an assignment, an over-approximation of all the possible locations $\abs{l}$ and of the dependencies $\abs{d}$ of the assigned expression $\jsexpr{e}$ is computed.
The abstract environment is then updated by setting the variable $\jsid{x}$ to this new abstract value.
All those possible flows from a potential dependency $\jsid{y} \in \abs{d}$ of the expression $\jsexpr{e}$ are marked as flowing into $\jsid{x}$.
The position of $\jsid{x}$ is taken into account in the resulting flows.

\begin{figure}
  \centering
  \begin{mathpar}
	\inferrule*[right=Skip]{ }{
		E^\sharp, H^\sharp, \jsexpr{\jscom{skip}} \absto E^\sharp, H^\sharp, \absset{}
	} \and
	\inferrule*[right=Seq]{
		E^\sharp, H^\sharp, \jsexpr{s_1} \absto E_1^\sharp, H_1^\sharp, \Delta^\sharp_1 \\
		E_1^\sharp, H_1^\sharp, \jsexpr{s_2} \absto E_2^\sharp, H_2^\sharp, \Delta^\sharp_2
	}{
		E^\sharp, H^\sharp, \jsexpr{s_1 ; s_2} \absto E_2^\sharp, H_2^\sharp, \Delta^\sharp_1 \cup \Delta^\sharp_2
	} \and
	\inferrule*[right=If]{
		E^\sharp, H^\sharp, \jsexpr{s_1} \absto E_1^\sharp, H_1^\sharp, \Delta^\sharp_1 \\
		E^\sharp, H^\sharp, \jsexpr{s_2} \absto E_2^\sharp, H_2^\sharp, \Delta^\sharp_2
	}{
		E^\sharp, H^\sharp, \jsif{e}{s_1}{s_2} \absto E_1^\sharp \sqcup E_2^\sharp, H_1^\sharp \sqcup H_2^\sharp, \Delta^\sharp_1 \cup \Delta^\sharp_2
	} \and
	\inferrule*[right=While]{
		E^\sharp \sqsubseteq E_0^\sharp \\
		H^\sharp \sqsubseteq H_0^\sharp \\
		E_0^\sharp, H_0^\sharp, \jsexpr{s} \absto E_1^\sharp, H_1^\sharp, \Delta^\sharp \\
		E_1^\sharp \sqsubseteq E_0^\sharp \\
		H_1^\sharp \sqsubseteq H_0^\sharp
	}{
		E^\sharp, H^\sharp, \jswhile{e}{s} \absto E_0^\sharp, H_0^\sharp, \Delta^\sharp
	} \and
	\inferrule*[right=Asg]{
		E^\sharp, H^\sharp, \jsexpr{e} \absto l^\sharp, d^\sharp
	}{
		E^\sharp, H^\sharp, \jsass[p]{x}{e} \absto E^\sharp[\jsid{x} \mapsto (l^\sharp, d^\sharp)], H^\sharp, \left(\abs{l} \cup d^\sharp\right) \abs\flowitem \jsid[p]{x}
	} \and
	\inferrule*[right=FldAsg]{
		E^\sharp, H^\sharp, \jsexpr{e_1} \absto l_1^\sharp, d_1^\sharp \\
		E^\sharp, H^\sharp, \jsexpr{e_2} \absto l_2^\sharp, d_2^\sharp
	}{
		E^\sharp, H^\sharp, \jsfieldass[p]{e_1}{f}{e_2} \absto E^\sharp, H^\sharp \sqcup \absheap{\left(l_1^\sharp, \jsid{f}\right) \mapsto \left(l_2^\sharp, d_2^\sharp\right)}, \left(l_2^\sharp \cup d_2^\sharp\right) \abs\flowitem l_1^\sharp\jscom{.}\jsid[p]{f}
	} \and
	\inferrule*[right=Del]{
		E^\sharp, H^\sharp, \jsexpr{e} \absto l^\sharp, d^\sharp \\
	}{
		E^\sharp, H^\sharp, \jsdelete{e}{f} \absto E^\sharp, H^\sharp, d^\sharp \abs\flowitem l^\sharp\jscom{.}\jsid{f}
	} \and
	\inferrule*[right=Cst]{ }{
		E^\sharp, H^\sharp, \jsexpr{c} \absto \absset{}, \absset{}
	} \and
	\inferrule*[right=Var]{
		E^\sharp[\jsid{x}] \sqsubseteq \left(l^\sharp, d^\sharp\right)
	}{
		E^\sharp, H^\sharp, \jsexpr{\jsid[p]{x}} \absto l^\sharp, \absset{\jsid[p]{x}} \cup d^\sharp
	} \and
	\inferrule*[right=Bin]{
		E^\sharp, H^\sharp, e_1 \absto l_1^\sharp, d_1^\sharp \\
		E^\sharp, H^\sharp, e_2 \absto l_2^\sharp, d_2^\sharp
	}{
		E^\sharp, H^\sharp, \jsop{e_1}{e_2} \absto l_1^\sharp\,op^\sharp\,l_2^\sharp, d_1^\sharp \cup d_2^\sharp
	} \and
	\inferrule*[right=Obj]{ }{
		E^\sharp, H^\sharp, \jsobj[p]{} \absto \absset{\objatpp{p}}, \absset{}
	} \and
	\inferrule*[right=Fld]{
		E^\sharp, H^\sharp, \jsexpr{e} \absto l^\sharp, d^\sharp \\
		H^\sharp[l^\sharp] \sqsubseteq o^\sharp \\
		o^\sharp[\jsid{f}] \sqsubseteq \left(l_f^\sharp, d_f^\sharp\right)
	}{
		E^\sharp, H^\sharp, \jsfield[p]{e}{f} \absto l_f^\sharp, \left(l^\sharp\jscom{.}\jsid[p]{f}\right) \cup d^\sharp \cup d_f^\sharp
	}
  \end{mathpar}
  \caption{Rules for the Abstract Reduction Relation}
  \label{fig:abstractreduction}
\end{figure}

The \textsc{Bin} rule makes use of an abstract operation $\abs{op}$, which
depends on the operators added in the language.
Most of the time, it shall be either $\sqcup$, or
the operation ignoring its operands and returning $\absset{}$.  For
instance, if the rules of $+$ forces its two operands to be integers,
raising an uncatchable error if one of them is an object, it's safe to
suppose $\abs{+}$ to be equal to $\lambda d_1^\sharp
d_2^\sharp. \absset{}$ as its returned value cannot be an object.
Figure~\ref{fig:abstractreductionexample} shows an example of analysis on the code we have seen on the previous sections, namely  $\jsexpr{\jsass{x}{\jsobj{}}; \jsfieldass{\jsid{x}}{f}{\jsobj{}}; \jsif{\jscom{false}}{\jsass{y}{\jsfield{\jsid{x}}{f}}}{\jsass{y}{\jsobj{}}}}$.

\begin{figure}
  \centering
  \begin{mathpar}
	E_1^\sharp = \absheap{\jsid{x} \mapsto \left(\absset{\objatpp{p_1}}, \absset{}\right)} \and
	E_2^\sharp = \absheap{\jsid{x} \mapsto \left(\absset{\objatpp{p_1}}, \absset{}\right), \jsid{y} \mapsto \left(\absset{\objatpp{p_2}}, \absset{}\right)} \and
	E_3^\sharp = \absheap{\jsid{x} \mapsto \left(\absset{\objatpp{p_1}}, \absset{}\right), \jsid{y} \mapsto \left(\absset{\objatpp{p_3}}, \absset{}\right)} \and
	E_4^\sharp = \absheap{\jsid{x} \mapsto \left(\absset{\objatpp{p_1}}, \absset{}\right), \jsid{y} \mapsto \left(\absset{\objatpp{p_2}, \objatpp{p_3}}, \absset{}\right)} \and
	H_1^\sharp = \absheap{\left(\objatpp{p_1}, \jsid{f}\right) \mapsto \absset{\objatpp{p_2}}, \absset{}}\\
    \inferrule*[Right=Seq]{
      \inferrule*[Left=Asg]{
        \inferrule*[Left=Obj]{ }
        {
          \bot, \bot, \jsexpr{\jsobj[p_1]{}} \absto \absset{\objatpp{p_1}}, \absset{}
        }
      }{
        \bot, \bot, \jsass{x}{\jsobj[p_1]{}} \absto E_1^\sharp, \bot, \absset{\objatpp{p_1} \flowitem \jsid{x}}
      }\\
      \inferrule*[Right=Seq,leftskip=27em,rightskip=10em,vdots=5em]{
        \inferrule*[Left=FldAsg]{
          \inferrule*[Left=Var]{
            E_1^\sharp[\jsid{x}] \sqsubseteq \left(\absset{\objatpp{p_1}}, \absset{}\right)
          }{
            E_1^\sharp, \bot, \jsid[p_2']{x} \absto \absset{\objatpp{p_1}}, \absset{\jsid[p_2']{x}}
          }\\
          \inferrule*[Right=Obj]{ }
          {
            E_1^\sharp, \bot, \jsexpr{\jsobj[p_2]{}} \absto \absset{\objatpp{p_2}}, \absset{}
          }
        }{
          E_1^\sharp, \bot, \jsfieldass{\jsid[p_2']{x}}{f}{\jsobj[p_2]{}}
		  \absto E_1^\sharp, H_1^\sharp, \absset{\objatpp{p_2} \flowitem \objatpp{p_1}\jscom{.}\jsid{f}}
        }\\
        \inferrule*[Right=If,leftskip=23em,rightskip=2em,vdots=6em]{
          \inferrule*[Left=Asg,vdots=4em,rightskip=3em]{
            \inferrule*[Left=Fld,rightskip=3em]{
              \inferrule*[Left=Var]{
                E_1^\sharp[\jsid{x}] \sqsubseteq \left(\absset{\objatpp{p_1}}, \absset{}\right)
              }{
                E_1^\sharp, H_1^\sharp, \jsexpr{\jsid{x}}
		        \absto \absset{\objatpp{p_1}}, \absset{}
              }\\
			  H_1^\sharp\left[\absset{\objatpp{p_1}}\right]\left[\jsid{f}\right] = \left(\absset{\objatpp{p_2}}, \absset{}\right)
			}{
              E_1^\sharp, H_1^\sharp, \jsfield{\jsid{x}}{f}
			  \absto \absset{\objatpp{p_2}}, \absset{}
			}
          }{
            E_1^\sharp, H_1^\sharp, \jsass{y}{\jsfield{\jsid{x}}{f}}
			\absto E_2^\sharp, H_1^\sharp, \absset{\objatpp{p_2} \flowitem \jsid{y}}
          }\\
          \inferrule*[Right=Asg]{
            \inferrule*[Right=Obj]{ }
            {
              E_1^\sharp, H_1^\sharp, \jsexpr{\jsobj[p_3]{}}
			  \absto \absset{\objatpp{p_3}}, \absset{}
            }
          }{
            E_1^\sharp, H_1^\sharp, \jsass{y}{\jsobj{}}
			\absto E_3^\sharp, H_1^\sharp, \absset{\objatpp{p_3} \flowitem \jsid{y}}
          }
        }
        {
          E_1^\sharp, H_1^\sharp, \jsif{\jscom{false}}{\jsass{y}{\jsfield{\jsid{x}}{f}}}{\jsass{y}{\jsobj{}}}
		  \absto E_4^\sharp, H_1^\sharp, \absset{\absset{\objatpp{p_2}, \objatpp{p_3}} \flowitem \jsid{y}}
        }
      }
      {
        E_1^\sharp, \bot, \jsexpr{\jsfieldass{\jsid{x}}{f}{\jsobj{}}; \jsif{\jscom{false}}{\jsass{y}{\jsfield{\jsid{x}}{f}}}{\jsass{y}{\jsobj{}}}}
        \absto E_4^\sharp, H_1^\sharp, \absset{\objatpp{p_2} \flowitem \objatpp{p_1}\jscom{.}\jsid{f}, \absset{\objatpp{p_2}, \objatpp{p_3}} \flowitem \jsid{y}}
      }
    }{
      \bot, \bot, \jsexpr{\jsass{x}{\jsobj[p_1]{}}; \jsfieldass{\jsid{x}}{f}{\jsobj[p_2]{}}; \jsif{\jscom{false}}{\jsass{y}{\jsfield{\jsid{x}}{f}}}{\jsass{y}{\jsobj[p_3]{}}}}
      \absto E_4^\sharp, H_1^\sharp, \absset{\objatpp{p_1} \flowitem \jsid{x}, \objatpp{p_2} \flowitem \objatpp{p_1}\jscom{.}\jsid{f}, \absset{\objatpp{p_2}, \objatpp{p_3}} \flowitem \jsid{y}}
    }
  \end{mathpar}
  \caption{Analysis Example}
  \label{fig:abstractreductionexample}
\end{figure}

There are several possible variations and extensions this
analysis. For one notable example 
it could be refined with strong updates on locations. For the moment,
we leave for further work how exactly to annotate the semantics and
to abstract locations in order to state whether or not an abstract
location represents a unique concrete location in the heap.

\subsection{Analysis in Coq}

The abstract domains are essentially the same as the ones described
in Section~\ref{sec:analysis:domains}.
They are straightforward to formalise as soon as basic constructions
for lattices are available:  the abstract domains are just specific
instances of standard lattices from abstract interpretation (flat
lattices, power set lattices\ldots).
For the certification of lattices we refer to the Coq developments by
David Pichardie~\cite{FICS08:Pichardie}. 

Similarly to Section~\ref{sec:annotatedsemanticincoq}, the rules of the analyser
presented in Figure~\ref{fig:abstractreduction} are first defined
as an inductive predicate of type
\begin{coq}
t AEnvironment -> t AHeap -> stat -> t AEnvironment
  -> t AHeap -> t AFlows -> Prop
\end{coq}
where the two types \coqinline|t AEnvironment| and \coqinline|t AHeap|
are the types of the abstract lattices for environments and heaps, and
\coqinline|t AFlows| the type of abstract flows, represented as a
lattice for convenience.
The analyser is then defined by an extractable function of similar
type (excepting the final ``\coqinline|-> Prop|''), the two
definitions being proven equivalent. 
The situation for expressions is similar.


Once the analysis has been defined as well as the instrumentation,
it's possible to formally prove the correctness of the abstract
reduction rules with respect to the instrumentation.
The property to prove is the one shown in
Section~\ref{sec:analysis:abstractreduction}:  if from an empty heap,
a program reduces to a heap $E_\tau$, $H_\tau$ and flows $\Delta_\tau$, then if from
the $\bot$ abstraction, a program reduces to $\abs{E}$, $\abs{H}$ and
abstract flows $\Delta^\sharp$; that is,
\begin{mathpar}
  \left[\right], \mset{}, [], \mset{}, \mset{}, \jsexpr{s} \to \tau, M_{\tau}, \Delta_\tau, E_{\tau}, H_{\tau}
  \and \mathrm{and} \and
  \bot, \bot, \jsexpr{s} \absto E^\sharp, H^\sharp, \Delta^\sharp
\end{mathpar}
then $E \prec \abs{E}$, $H \prec \abs{H}$ and $\Delta_\tau \prec \Delta^\sharp$.
This can be followed on \cite{OWhileFlowsSourceCode}.

\section{Conclusion}

Schmidt's natural semantics-based abstract interpretation is a rich
framework which can be instantiated in a number of ways. In this
paper, we have shown how the framework can be applied to the
particular style of natural semantics called pretty big step
semantics. We have studied a particular kind of intensional
information about the program execution, \emph{viz.}, how information
flows from points of creation to points of use. This has lead us to
define a particular abstraction of semantic derivation trees for
describing points in the execution.  This abstraction can then be
further combined with other abstractions to obtain an abstract
reduction relation that formalizes the static analysis. 

Other systematic derivation of static analyses have
taken small-step operational semantics as starting
point. Cousot~\cite{Cousot:98:Marktoberdorf} has shown how to
systematically derive static analyses for an imperative
language using the principles of abstract interpretation. Midtgaard
and Jensen\cite{MidtgaardJ:08,MidtgaardJ:09} used a similar
approach for calculating control-flow analyses for functional languages
from operational semantics in the form of abstract machines. Van Horn
and Might~\cite{VanHorn:10:Abstracting} show how a series of
analyses for functional languages can be derived from abstract
machines. An advantage of using small-step semantics
is that the abstract interpretation theory is conceptually simpler and
more developed than its big-step counterpart. Our motivation for
developing the big-step approach further is that the semantic
framework has certain modularity properties that makes it a popular
choice for formalizing real-sized programming languages. 

The semantics and its abstractions lend themselves well to being
implemented in the proof assistant Coq. This is an important point, as
some form of mechanization is required to evaluate the scalability of
the method.  Scalability is indeed one of the goals for this work. The
present paper establishes the principles with which we hope to achieve
the generation of an analysis for full \javascript based on its Coq
formalization. However, this will require some form of
machine-assistance in the production of the abstract semantics. The
present work provides a first experience of how to proceed. Further
work will now have to extract the essence of this process and
investigate how to program it in Coq.

One this has been achieved, we will be well armed to attack other
analyses. One immediate candidate for further work is full information
flow analysis, taking indirect flows due to conditionals into
account. It would in particular be interesting to see if the resulting
abstract semantics can be used for a rational reconstruction of the
semantic foundations underlying the 
dynamic and hybrid information flow analysis techniques developed by
Le Guernic, Banerjee, Schmidt and Jensen \cite{LeGu-etal-06-ASIAN}. Combined with the
extension to full \javascript, this would provide a certified version
of the recent information flow control mechanisms for \javascript such
as the monitor proposed by Hedin and
Sabelfeld \cite{Hedi-Sabe-12-CSF}. 

We hope to report on this in the next Festschrift to David Schmidt.

\begin{figure}
  \centering
  \begin{align*}
    \appo{Name}{\ppn} &= \ppb{Name}\ppn && \text{for normal names}\\
    \appo{Name}{\ppn} &= \ppn && \text{for extended names}\\
    \appi{Seq}{\ppn} &= \ppb{Seq1}\ppn\\
    \appi{Seq1}{\ppn} &= \ppb{Seq2}\ppn\\
    \appi{If}{\ppn} &= \ppb{IfE}\ppn\\
    \appi{IfTrue}{\ppn} &= \ppb{If1}\ppn\\
    \appi{IfFalse}{\ppn} &= \ppb{If2}\ppn\\
    \appi{While}{\ppn} &= \ppb{WhileE}\ppn\\
    \appi{WhileTrue1}{\ppn} &= \ppb{WhileS}\ppn\\
    \appi{Asg}{\ppn} &= \ppb{AsgE}\ppn\\
    \appi{FldAsg}{\ppn} &= \ppb{FldAsg1}\ppn\\
    \appi{FldAsg1}{\ppn} &= \ppb{FldAsg2}\ppn\\
    \appi{Del}{\ppn} &= \ppb{DelE}\ppn\\
    \appi{Bin}{\ppn} &= \ppb{Bin1}\ppn\\
    \appi{Bin1}{\ppn} &= \ppb{Bin2}\ppn\\
    \appi{Fld}{\ppn} &= \ppb{FldE}\ppn\\
    \appi{Name}{\ppn} &= \ppn && \text{otherwise}\\
    \pusho{Name} &= \orule{Name} :: [] && \text{for normal names}\\
    \pusho{Name} &= [] && \text{for extended names}\\
    \pushi{Seq1} &= \orule{Seq} :: []\\
    \pushi{IfTrue} &= \orule{If} :: []\\
    \pushi{IfFalse} &= \orule{If} :: []\\
    \pushi{WhileTrue1} &= \orule{While} :: []\\
    \pushi{WhileTrue2} &= \orule{While} :: []\\
    \pushi{Asg1} &= \orule{App} :: []\\
    \pushi{FldAsg1} &= \orule{FldAsg} :: []\\
    \pushi{FldAsg2} &= \orule{FldAsg} :: []\\
    \pushi{Bin1} &= \orule{Bin} :: []\\
    \pushi{Bin2} &= \orule{Bin} :: []\\
    \pushi{Del1} &= \orule{Del} :: []\\
    \pushi{Fld1} &= \orule{Fld} :: []\\
    \pushi{Name} &= [] && \text{otherwise}
\end{align*}
  
  \caption{Helper functions to convert traces to program points}
  \label{fig:tracespphelper}
\end{figure}

\bibliographystyle{eptcs}

\end{document}

